\pdfoutput=1
\documentclass[letterpaper,titlepage,11pt]{article}

\usepackage{jheppub}
\usepackage[T1]{fontenc}
\usepackage[latin9]{inputenc}
\usepackage[active]{srcltx}
\usepackage{textcomp}
\usepackage{amsmath}
\usepackage{amssymb}
\usepackage{esint}
\usepackage{multirow}

\usepackage{textcomp}

\newcommand{\eq}[2]{\begin{equation} #1 \label{#2} \end{equation}}

\def\cC{\mathcal{C}}

\def\cG{\mathcal{G}}

\def\cL{\mathcal{L}}
\def\cM{\mathcal{M}}
\def\cN{\mathcal{N}}
\def\cO{\mathcal{O}}
\def\cP{\mathcal{P}}

\usepackage{enumerate}

\def\be{\begin{eqnarray}}
\def\ee{\end{eqnarray}}
\def\beann{\begin{eqnarray*}}
\def\eeann{\end{eqnarray*}}
\def\beq{\begin{equation}}
\def\eeq{\end{equation}}
\def\ba{\begin{array}}
\def\ea{\end{array}}
\def\ben{\begin{enumerate}}
\def\een{\end{enumerate}}
\def\bea{\begin{eqnarray}}
\def\eea{\end{eqnarray}}

\def\eps{\epsilon}
\def\ve{\varepsilon}

\renewcommand{\d}{\partial}

\usepackage{amsfonts}

\usepackage{tikz}

\usepackage{bbm}

\DeclareMathOperator{\extdm}{d}
\newcommand{\extd}{\extdm \!}

\title{Boundary theories for dilaton supergravity in 2D}

\subheader{TUW-18-05}

\newcommand{\marcela}{\ast}
\newcommand{\oscar}{\dagger}
\newcommand{\hernan}{\ddagger}
\newcommand{\daniel}{\S}
\newcommand{\carlos}{\P}
\newcommand{\dima}{\parallel}

\author[\marcela]{Marcela C\'ardenas,}
\emailAdd{cardenas@apc.in2p3.fr}
\author[\oscar]{Oscar Fuentealba,}
\emailAdd{fuentealba@cecs.cl}
\author[\hernan]{Hern\'an A. Gonz\'alez,}
\emailAdd{hgonzale@hep.itp.tuwien.ac.at}
\author[\daniel]{Daniel Grumiller,}
\emailAdd{grumil@hep.itp.tuwien.ac.at}
\author[\carlos]{Carlos Valc\'arcel}
\emailAdd{valcarcel.flores@gmail.com}
\author[\dima]{and Dmitri Vassilevich}
\emailAdd{dvassil@gmail.com}

\affiliation[\marcela]{APC, Universit\'e Paris Diderot \&
CNRS, CEA, Observatoire de Paris \\
Sorbonne Paris Cit\'e 10, rue Alice Domon et L\'eonie Duquet, F-75205 Paris CEDEX 13, France}

\affiliation[\marcela]{Universidad Cat\'olica del Maule, Av. San Miguel 3605, Talca, Chile}
\affiliation[\marcela,\oscar,\daniel,\hernan]{Erwin-Schr\"odinger International Institute for Mathematics and Physics\\
Boltzmanngasse 9A, A-1090 Vienna, Austria}

\affiliation[\oscar,\daniel]{Centro de Estudios Cient\'ificos (CECs), Av. Arturo Prat 514, Valdivia, Chile}
\affiliation[\marcela,\oscar,\hernan,\daniel]{Institute for Theoretical Physics, TU Wien, Wiedner Hauptstr.~8-10/136, A-1040 Vienna, Austria}
\affiliation[\carlos]{Instituto de F\'isica - Universidade Federal da Bahia, C\^ampus Universit\'ario de Ondina, 40210-340, Salvador, B.A. Brazil}
\affiliation[\carlos,\dima]{CMCC-Universidade Federal do ABC, Santo Andr\'e, S.P. Brazil}
\affiliation[\dima]{Department of Physics, Tomsk State University, Tomsk, Russia}

\abstract{The $\mathfrak{osp}(2,N)$--BF formulation of dilaton supergravity in two dimensions is considered. We introduce a consistent class of asymptotic conditions preserved by the extended superreparametrization group of the thermal circle at infinity. In the $N=1$ and $N=2$ cases the phase space foliation in terms of orbits of the super-Virasoro group allows to formulate suitable integrability conditions for the boundary terms that render the variational principle well-defined. Once regularity conditions are imposed, requiring trivial holonomy around the contractible cycle the asymptotic symmetries are broken to some subsets of exact isometries. Different coadjoint orbits of the asymptotic symmetry group yield different types of boundary dynamics; we find that the action principle can be reduced to either the extended super-Schwarzian theory, consistent with the dynamics of a non-vanishing Casimir function, or to superparticle models,  compatible with bulk configurations whose Casimir is zero. These results are generalized to $\cN \geq 3$ by making use of boundary conditions consistent with the loop group of OSp$(2,N)$. Appropriate integrability conditions permit to reduce the dynamics of dilaton supergravity to a particle moving on the OSp$(2,N)$ group manifold. Generalizations of the boundary dynamics for $\cN>2$ are obtained once bulk geometries are supplemented with super-AdS$_2$ asymptotics.}

\begin{document}

\maketitle 


\section{Introduction}

Gauge symmetry is generically linked to the redundancy in the variables used to describe a field theory. This association is nevertheless not entirely true in theories with boundaries: part of the gauge degrees of freedom becomes dynamical on these boundaries. The working example of this realization is the fact that three-dimensional Chern--Simons theories in presence of a boundary can be expressed as a WZW model in two dimensions \cite{Witten:1988hf,Elitzur:1989nr}. Since then, this relation has been extended to topological theories of gravity \cite{Balachandran:1994vi,Coussaert:1995zp,Bergamin:2005pg,Barnich:2013yka,Carlip:2016lnw,Maldacena:2016upp,Geiller:2017xad} and more recently to theories with local degrees of freedom \cite{Donnelly:2015hxa,Donnelly:2016auv,Blommaert:2018oue,Barnich:2018zdg}. Deepening further this program can help to understand the role played by boundary modes in the context of quantum gravity. From a more general perspective, this can shed light on the origin of the holographic behavior of gravity \cite{tHooft:1993dmi,Susskind:1994vu}.
 

Here we focus on the study of boundary dynamics associated with two-dimensional BF theories based on the gauge group OSp$(2,N)$. These lagrangians are supersymmetric extensions of the so-called Jackiw--Teitelboim (JT) model \cite{Jackiw:1984, Teitelboim:1983ux}.  
The motivation for this is twofold. On the one hand, the addition of local supersymmetry is a natural way to generalize the results of \cite{Gonzalez:2018enk}, where the asymptotic dynamics was analyzed for (bosonic) higher spin extensions of the JT theory. On the other hand, it is known that supersymmetric BF models provide gravitational duals for supersymmetric extensions of the SYK model in the low temperature regime \cite{Fu:2016vas,Peng:2018zap,Bulycheva:2018qcp,Narayan:2017hvh}. In this sense, the main purpose of this paper is to explore the asymptotic dynamics of two-dimensional dilaton supergravity and study its connection with models placed at the conformal boundary of the spacetime; namely the super-Schwarzian theory and superconformal quantum mechanics \cite{deAlfaro:1976je,Akulov:1984uh,Fubini:1984hf}. This analysis yields a consistent setup that extends and complements previous analyses in the $N=1$ case \cite{Astorino:2002bj} and more recent ones in the second order formulation of dilaton supergravity \cite{Forste:2017apw,Forste:2017kwy} for $\cN=1$ and $\cN=2$ supersymmetries. 


Working with BF models allows to introduce a generalized notion of Euclidean black holes: they can be regarded as gauge fields with trivial holonomies along the thermal cycle. In two dimensions, this condition adds restrictions on the set of configurations with given asymptotics. Specifically, restricting the phase space to globally well-defined black holes produces an explicit 
breaking of the asymptotic symmetry group. Notably, this effect has been observed in dual descriptions controlled by the  Schwarzian theory \cite{Kitaev:15ur,Maldacena:2016hyu,Maldacena:2016upp}. 

 
 In this work, we propose two different sets of boundary conditions with fluctuating dilaton multiplets. The first one reproduces the SO$(N)$--extended superreparametrization group around the thermal circle at infinity, SDiff$(S^{1})$. The second one is preserved by the loop group of OSp$(2,N)$. For the former set of asymptotic conditions, the reduced action principle can be determined by two different boundary models related to specific coadjoint orbits of SDiff$(S^{1})$. Concretely, we show that the asymptotic dynamics described by the extended super-Schwarzian theory is consistent with a nonvanishing Casimir function while the dynamics compatible with zero Casimir entails the existence of a supersymmetric particle model at the boundary. In both cases, however, the phase space is written in terms of orbits of the super-Virasoro group, which is only accessible for $N=1$ and $N=2$. This is due to quadratic contributions of affine $\mathfrak{so}(N)$ currents in the infinitesimal transformation laws $N\geq 3$.\footnote{This kind of non-linearities also appears in the case of extended AdS supergravity in 3D \cite{Henneaux:1999ib}.} In order to access the asymptotic dynamics for $N\geq 3$,  we make use of the second set of boundary conditions, consistent with the loop group of OSp$(2,N)$, which allows to reformulate the problem in terms of a particle moving on the group manifold. Further constraints on the latter reduced model are introduced when requiring to have asymptotically super-AdS$_2$ solutions in the bulk. This procedure gives rise to supersymmetric extensions of particle models without self-interacting potentials and extended super-Schwarzian theories with $\cN>2$.


The paper is organized as follows. In the next section, we present the BF formulation of minimal dilaton supergravity in two dimensions with gauge group OSp$(2,1)$. In section \ref{sec:Asymptotic conditions}, a consistent set of asymptotic conditions is proposed together with a superspace formulation that describes in a super-covariant way the action of the asymptotic symmetries on the fields. The latter approach turns out to be particularly useful in order to foliate the phase space of the theory in terms of the adjoint and coadjoint actions of the super-Virasoro group. Using this prescription, we demand suitable boundary conditions that render the variational principle well--defined. This is discussed in section \ref{biendef}. Section \ref{sec:Holonomy} is devoted to analyze regularity conditions on solutions describing Euclidean black holes. It is shown that configurations with trivial holonomies are solutions of the supersymmetric Hill equation, which must obey (anti)periodic boundary conditions. In Section \ref{bdf}, the reduced action is obtained by evaluating the variational principle on the constraint surface, which for our set of asymptotic conditions makes the bulk contribution vanish. Therefore, the regularized action is only determined by a boundary term that leads -- after restricting to regular solutions -- to either the $\cN=1$ super-Schwarzian theory, or to a supersymmetric particle model. In section \ref{extended}, we present asymptotic conditions extending the previous results but mainly focusing on the $\cN=2$ case, arriving at analogous conclusions. Section \ref{ultimasec} deals with the boundary dynamics of $\cN>2$ extended dilaton supergravity, where supersymmetric extensions of conformal mechanics and Schwarzian theories are found. In section \ref{se:9} we go beyond the highest-weight ansatz to consider more general supersymmetric asymptotically AdS$_2$ boundary conditions. 
Finally,  section \ref{CR} is dedicated to discuss possible applications and extensions of this work. Conventions associated to the explicit representation of $\mathfrak{osp}(2,N)$ generators can be found in appendix \ref{sec:A1}.

Before starting we mention that we work in Euclidean signature throughout the whole paper. Thus, when speaking about ``time'' we always mean ``Euclidean time''.

\section{\texorpdfstring{$\boldsymbol{\mathcal{N}=1}$}{N=1} dilaton supergravity in two dimensions}\label{sec:Dilaton supergravity theory}

The first order action for the minimal supersymmetric extension of the JT 
model \cite{Teitelboim:1983uy,Chamseddine:1991fg} reads\footnote{The form of the action implies that we have adopted the convention $(\psi_1 \psi_2)^*=\psi_1 \psi_2$ for the complex conjugate of the product of two real Grassmann variables.} 

\begin{equation}
I=\frac{k}{4\pi}\intop_\cM \big[ X^{a}\big(\extd e_{a}+\epsilon_{ab}\omega e^{b}-\tfrac{1}{4}\bar{\psi}\gamma_{a}\psi\big) + X\extd\omega + \tfrac{1}{2}X\big(\epsilon_{ab}e^{a}e^{b}+\tfrac{1}{2}\bar{\psi}\gamma_{5}\psi\big) + \bar{\lambda}D\psi+\tfrac{1}{2}\bar{\lambda}e^{a}\gamma_{a}\psi\big]\label{eq:SJTaction}
\end{equation}
where the manifold $\cM$ has the topology of a disk, endowed with a time $\tau$ identified as $\tau \sim \tau +\beta$ and a non-compact direction $r>0$. 

The field content of this theory is the following: $X^{a}$ is a pair of Lagrange
multipliers enforcing the supercovariant torsion constraint, $X$ stands
for the dilaton field, $\lambda^{\alpha}$ is the dilatino field,
$e^{a}$ corresponds to the one-form zweibein, $\omega$ is the dualized spin connection and $\psi^{\alpha}$
is the gravitino one-form. The covariant derivative acting on the gravitino is
defined as
\begin{equation}
D\psi=\extd\psi+\frac{1}{2}\omega\gamma_{5}\psi\,.
\end{equation}
The indices are contracted with the flat metric $\delta_{ab}$
and $\epsilon_{01}=-1$, where $a,b=0,1$. The Majorana conjugate
is defined as $\bar{\psi}_{\alpha}=\psi^{\beta}C_{\beta\alpha}$,
where $C_{\alpha\beta}=\epsilon_{\alpha\beta}$ ($\epsilon_{+-}=-1$)
is the charge conjugation matrix. The matrices $\gamma_{a}$ satisfy
the two-dimensional Clifford algebra, $\{\gamma_{a},\gamma_{b}\}=2\delta_{ab}$,
and $\gamma_{5}=-\gamma_{0}\gamma_{1}$, then $\gamma_{5}^{2}=-\mathbb{I}_{2}$.

It is possible to show that action \eqref{eq:SJTaction}, apart from being invariant under local Lorentz transformations and conformal boosts, is invariant under the following local supersymmetry transformations
\begin{align}
\delta e^{a}&=-\frac{1}{2}\bar{\psi}\gamma^{a}\epsilon & \delta\omega&=\frac{1}{2}\bar{\psi}\gamma_{5}\epsilon & \delta\psi& =D\epsilon+\frac{1}{2}e^{a}\gamma_{a}\epsilon\label{eq:d1}\\
\delta X^{a}&=-\frac{1}{2}\bar{\lambda}\gamma^{a}\epsilon & \delta X&=-\frac{1}{2}\bar{\lambda}\gamma_{5}\epsilon & \delta\lambda&=\frac{1}{2}X^{a}\gamma_{a}\epsilon-\frac{1}{2}X\gamma_{5}\epsilon\,.\label{eq:d2}
\end{align}

Minimal dilaton supergravity \eqref{eq:SJTaction} does not propagate local degrees of freedom. This property is made manifest by writing the action in terms of a BF-theory for OSp$(2,1)$ \cite{Montano:1990ru}. Its bulk action reads
\begin{equation}
I=\frac{k}{2\pi}\intop_{\cM} \text{Str}\left[\textbf{X}\left(\extd\textbf{A}+\textbf{A}^{2}\right)\right]\label{eq:BFaction}
\end{equation}
where the zero-form dilaton multiplet $\textbf{X}=X^{I}J_{I}+\lambda^{\alpha}Q_{\alpha}$
and the one-form connection $\textbf{A}=A^{I}J_{I}+\psi^{\alpha}Q_{\alpha}$,
are valued in the $\mathfrak{osp}(2,1)$ algebra
\begin{equation}
[J_{I},J_{J}]=\epsilon_{IJK}J^{K}\qquad\qquad[J_{I},Q_{\alpha}]=\frac{1}{2}(\Gamma_{I})_{\,\,\,\alpha}^{\beta}Q_{\beta}\qquad\qquad\{Q_{\alpha},Q_{\beta}\}=-\frac{1}{2}(C\Gamma^{I})_{\alpha\beta}J_{I}\label{eq:osp(1,2)}.
\end{equation}
Here, the $\Gamma$-matrices $\Gamma_{I}=\left(\gamma_{a},-\gamma_{5}\right)$,
satisfy the Clifford algebra in three dimensions, $\{\Gamma_{I},\Gamma_{J}\}=2\eta_{IJ}$,
where the metric is given by $\eta_{IJ}=\text{diag}(1,1,-1)$ and
the orientation is chosen as $\epsilon_{012}=-1$. The components
that go along the $\mathfrak{so}(2,1)$ generators collect the bosonic content
of the model, such that $X^{I}=(X^{a},X)$ and $A_{I}=(e_{a},\omega)$.
The nonvanishing components of the supertrace are\footnote{Explicit matrix representations for the $\mathfrak{osp}(2,N)$ generators are provided in appendix \ref{sec:A1}.}
\begin{equation}
\text{Str}[J_{I}J_{J}]=\frac{1}{2}\eta_{IJ}\qquad\qquad\text{Str}[Q_{\alpha},Q_{\beta}]=\frac{1}{2}C_{\alpha\beta}\,.
\end{equation}

Let us review some of the features associated with the action \eqref{eq:BFaction}. The field equations for the BF-theory generically read as follows
\begin{equation}
\textbf{F}=0\qquad\qquad \extd\textbf{X}+[\textbf{A},\textbf{X}]=0\,.\label{eq:BFeqns}
\end{equation}
In the case we are dealing with, the field strength $\textbf{F}=\extd\textbf{A}+\textbf{A}^{2}$ associated
to $\textbf{A}$, is explicitly given by
\begin{equation}
\textbf{F}=\big(R^{I}-\frac{1}{4}\bar{\psi}\Gamma^{I}\psi\big)\, J_{I}+\nabla\psi^{\alpha}Q_{\alpha}\label{eq:F}
\end{equation}
where the $\mathfrak{so}(2,1)$ curvature two-form $R^{I}$ and the covariant 
derivative $\nabla$ are, respectively,
\begin{equation}
R^{I} =  \extd A^{I}+\frac{1}{2}\epsilon^{IJK}A_{J}A_{K}\qquad\qquad \nabla\psi =  d\psi+\frac{1}{2}A^{I}\Gamma_{I}\psi\,. \label{Rypsi}
\end{equation}

A crucial feature of formulating the theory as \eqref{eq:BFaction} is that the gauge invariance of the model becomes manifest. Indeed, it is straightforward to prove that under the action of the gauge group, $\textbf{A}$ and $\textbf{X}$  transform in the adjoint and coadjoint representations, respectively, which leaves \eqref{eq:BFaction} unchanged. The form of the infinitesimal gauge transformations, generated by a Lie-algebra-parameter $\Lambda$, then read
\begin{equation}
\delta_{\Lambda}\textbf{A}=\extd\Lambda+[\textbf{A},\Lambda]\qquad\qquad\delta_{\Lambda} \textbf{X}=[\textbf{X},\Lambda]\,.\label{eq:Gaugetrafos}
\end{equation}
From the transformation rule of the gauge field in \eqref{eq:Gaugetrafos}, it is possible to conclude that the dilaton equation of motion can be written as  $\delta_{\textbf{X}}\textbf{A}\big|_{\rm on-shell}=0$. Therefore, the on-shell value of the dilaton $\textbf{X}$ turns out to be the stabilizer of the gauge field $\textbf{A}$ .

It is important to mention that  diffeomorphisms are a distinctive class of gauge symmetries. These are spanned by $\Lambda=\xi^{\sigma}\textbf{A}_{\sigma}$, whose infinitesimal group action \eqref{eq:Gaugetrafos} is expressed as Lie derivatives acting on $\textbf{A}_{\mu}$ and $\textbf{X}$, up to transformations of the form
\be
\label{trivial}
\delta_{\rm trivial}\textbf{A}_{\mu}= \chi_\mu \frac{\delta I}{\delta \textbf{X} }\qquad \qquad
\delta_{\rm trivial} \textbf{X} = -\chi_\mu \frac{\delta I}{\delta \textbf{A}_{\mu} }
\ee
where $\chi_{\mu}=\frac{2\pi}{k} \xi^{\sigma}\epsilon_{\sigma \mu}$. The latter type of transformations are indeed trivial symmetries of the action, no matter what parameter $\chi_{\mu}$ is being used (see, e.g., \cite{Henneaux:1992ig}).

It is reassuring to verify that gauge transformations \eqref{eq:Gaugetrafos}  generated by $\Lambda=\epsilon^{\alpha}Q_{\alpha}$,
\be
\delta_{\epsilon}A^{I}=-\frac{1}{2}\bar{\psi}\Gamma^{I}\epsilon\quad\qquad\delta_{\epsilon}X^{I}=-\frac{1}{2}\bar{\lambda}\Gamma^{I}\epsilon\quad\qquad \delta_{\epsilon}\psi=\nabla\epsilon\quad\qquad \delta_{\epsilon}\lambda=\frac{1}{2}X^{I}\Gamma_{I}\epsilon\label{eq:d2-1}
\ee
lead to the ones given in \eqref{eq:d1} and \eqref{eq:d2}.

Finally, we note that BF models admit the definition of a conserved quantitiy called (super) Casimir function, which is defined as
\be
\label{cas}
\cC\equiv-\frac{1}{2}\,\text{Str}\big(\textbf{X}^2\big)\,.
\ee
This is a gauge invariant object that satisfies the conservation equation $\extd\cC=0$ by virtue of the dilaton equation of motion $\delta_{\textbf{X}}\textbf{A}\big|_{\rm on-shell}=0$. 

\section{Asymptotic structure}\label{sec:Asymptotic conditions}

We are interested in studying the set of superdiffeomorphisms preserving
 the form of the fall-off of the fields at infinity, which corresponds to a subset of gauge transformations modulo trivial terms proportional 
to the equations of motion \eqref{trivial}. More precisely, we
proceed to find the symmetries of a class of configurations that asymptotically satisfies the conditions  
\be
\delta_{\Lambda} \textbf{A}_{\mu}= \cO(\textbf{A}_{\mu})\qquad \qquad
\delta_{\Lambda} \textbf{X}= \cO(\textbf{X})
\ee
where $\Lambda$ is the Lie-superalgebra-parameter that generates the infinitesimal transformations associated to the asymptotic symmetries.

\subsection{Fall-off conditions}\label{sec:fo}

Here, we present a set of boundary conditions for the theory, that besides of containing all the bosonic solutions of interest with non-trivial dilaton \cite{Grumiller:2015vaa,Grumiller:2016dbn,Grumiller:2017qao}, turns out to be preserved by SDiff$(S^{1})$ at infinity. The fall-off of the fields $\textbf{A}$ and $\textbf{X}$, written in the Coussaert-Henneaux-van Driel gauge \cite{Coussaert:1995zp}, reads as follows
\be
\textbf{A}[\cL,\psi]=b^{-1}(\extd+a)b+\cO(r^{-2}) \qquad\qquad \textbf{X}[y,\rho]=b^{-1}x b +\cO(r^{-2}) \label{AXExt1}
\ee
where $b=\exp[\log(r)L_0]$. The radially independent gauge field $a$ reads
\begin{align}
a&=\left[L_{1}+\cL L_{-1}+\mathcal{\psi}G_{-\frac{1}{2}}\right] \extd\tau \label{eq:A}
\end{align}
and the dilaton $x$ is given by
\be
x=y L_{1}-y'L_{0}+\big(\frac{1}{2}y\text{\ensuremath{''}}+y\cL+\psi \rho\big)L_{-1}+\rho G_{\frac{1}{2}}-\left(\rho\text{\ensuremath{'}}-y\mathcal{\psi}\right)G_{-\frac{1}{2}}\,. \label{eq:X}
\ee
We have expressed $\mathfrak{osp}(2,1)$ in terms of
the basis $\{L_{-1},L_{0},L_{1},G_{-\frac{1}{2}},G_{\frac{1}{2}}\}$,\footnote{For the change of basis see appendix \ref{sec:A1}.} such that
\be
\label{algebraLG}
\left[L_{m},L_{n}\right]=\left(m-n\right)L_{m+n}\;\qquad\left[L_{m},G_{p}\right]=\big(\frac{m}{2}-p\big)G_{m+p}\;\qquad\left\{ G_{p},G_{q}\right\} =-2L_{p+q}
\ee
with $m,n=\pm1,0$ and $p,q=\pm\frac{1}{2}$. The dynamical fields $\mathcal{L}$
and $y$ are arbitrary functions of time $\tau$, while
$\psi$ and $\rho$ correspond to arbitrary Grassmann-valued functions
of $\tau$. 

The gauge parameter that maintains the asymptotic form
of the configurations \eqref{eq:A}, \eqref{eq:X}  is given by
\begin{equation}
\Lambda=\mathbf{X}[\xi,\epsilon,\mathcal{L},\psi]\label{eq:Lambda}
\end{equation}
where $\xi$ and $\epsilon$ represent superdiffeomorphisms in the
thermal circle at infinity. Under these asymptotic symmetries, the
fields transform as
\begin{align}
\delta_{(\xi,\epsilon)}\mathcal{L} & =  2\mathcal{L}\xi\text{\ensuremath{'}}+\mathcal{L}\text{\ensuremath{'}}\xi+\frac{1}{2}\xi\text{\ensuremath{'''}}+3\psi\epsilon\text{\ensuremath{'}}+\psi\text{\ensuremath{'}}\epsilon\label{eq:deltaL}\\
\delta_{(\xi,\epsilon)}\psi & =  \frac{3}{2}\psi\xi\text{\ensuremath{'}}+\psi\text{\ensuremath{'}}\xi-\mathcal{L}\epsilon-\epsilon\text{\ensuremath{''}}\label{eq:deltapsi}\\
\delta_{(\xi,\epsilon)}y & =  \xi y\text{\ensuremath{'}}-\xi\text{\ensuremath{'}}y+2\epsilon\rho\label{eq:delta y}\\
\delta_{(\xi,\epsilon)}\rho & =  \frac{1}{2}\epsilon y\text{\ensuremath{'}}-\epsilon\text{\ensuremath{'}}y-\frac{1}{2}\xi\text{\ensuremath{'}}\rho+\xi\rho\text{\ensuremath{'}}\,.\label{eq:delta eps}
\end{align}
Note that equations \eqref{eq:deltaL} and \eqref{eq:deltapsi} correspond
to infinitesimal transformations of a super-Virasoro coadjoint orbit
with representatives $\mathcal{L}$ and $\psi$, while \eqref{eq:delta y} and \eqref{eq:delta eps} reflect that $y$ and $\rho$ transform in the adjoint representation of the super-Virasoro group \cite{Aoyama:1989pw,Aratyn:1989qq,Delius:1990pt}.

It is worth mentioning that the asymptotic form of $\textbf{X}$, given in \eqref{eq:X}, allows to incorporate fluctuating dilaton and dilatino without spoiling the superreparametrization group at infinity. This is because \eqref{eq:X} was chosen to have the same form of the gauge parameter \eqref{eq:Lambda} that in turn preserves the asymptotic behaviour of the gauge connection \eqref{eq:A}. This fact is reminiscent of the procedure introduced in \cite{Henneaux:2013dra}, that in the context of three-dimensional higher spin gravity permits to include a generic choice of chemical potentials without breaking the $W$-symmetry at infinity.

\subsection{Asymptotic field equations}\label{se:9.4}

The field equations for the gauge field
are asymptotically satisfied for the set of asymptotic conditions \eqref{AXExt1},
i.e., the strength field associated to the gauge field \eqref{eq:A}
vanishes at infinity
\be
\label{fe0}
\mathbf{F}=\mathcal{O}(r^{-2})\,.
\ee
The remaining equations of motion dictate that
the dilaton and dilatino acquire asymptotically the role of stabilizers
of the dynamical fields $\mathcal{L}$ and $\psi$, namely 
\be
\delta_{(y,\rho)}\mathcal{L}=0\qquad \qquad \delta_{(y,\rho)}\mathcal{\psi}=0\,.
\ee
By virtue of these two equations of motion, the \emph{off-shell} Casimir function \eqref{cas}
\be
\label{supercas}
 \cC=\mathcal{L}y^{2}+\frac{1}{2}yy''-\frac{1}{4}\left(y'\right)^{2} +3y\psi\rho +2\rho \rho'+\mathcal{O}(r^{-1})
\ee
defines an asymptotically conserved quantity, i.e.,  $\d_{\tau} \cC=\mathcal{O}(r^{-1})$.

\subsection{Superspace formulation}\label{se:9.5}

For later purposes we will describe the realization of the asymptotic symmetry group on the dynamical variables in the superspace formalism (see, e.g.,~\cite{Friedan:1986rx} and references therein). Superspace is defined by a pair of coordinates $(\tau,\theta)$ where $\theta$ denotes a Grassmann variable.  Superdiffeomorphisms correspond to coordinate transformations $\tilde{\theta}=\tilde{\theta}(\tau,\theta)$ and $\tilde{\tau}=\tilde{\tau}(\tau,\theta)$ that preserve the constraint
\be
\label{conssup}
D\tilde{\tau}=\tilde{\theta}D\tilde{\theta}
\ee
where the supercovariant derivative $D\equiv\partial_{\theta}+\theta\partial_{\tau}$ must transform as $D=D\tilde{\theta}\tilde{D}$. A parametrization that infinitesimally solves restriction \eqref{conssup}  is given by 
\be
\label{infpa}
\tilde{\tau}=\tau-\xi(\tau)+\theta\epsilon(\tau)\qquad \qquad  \tilde{\theta}=\theta+\epsilon(\tau)-\frac{1}{2}\theta\xi\text{\ensuremath{'}}(\tau)\,.
\ee

Let us translate the field content of the asymptotic conditions \eqref{AXExt1} into superfields. Quantities  $\mathcal L$ and $\psi$ can be combined into a single superfield $\Psi$ defined as
\begin{equation}
\Psi(\tau,\theta)\equiv\psi(\tau)+\theta\mathcal{L}(\tau)\,,\label{eq:def-Psi}
\end{equation}
which 
has conformal weight $\frac{3}{2}$.  The finite transformation of $\Psi$ under superdiffeormorphism $\tilde{\theta}$  is given by
\begin{equation}
\tilde{\Psi}(\tilde{\tau},\tilde{\theta})=(D\tilde{\theta})^{-3}\left[\Psi(\tau,\theta)-S(\tilde{\tau},\tilde{\theta};\tau,\theta) \right]\label{eq:barPsi}
\end{equation}
with $S(\tilde{\tau},\tilde{\theta};\tau,\theta)$ the
super-Schwarzian \cite{Cohn:1986wn}
\begin{equation}
S(\tilde{\tau},\tilde{\theta};\tau,\theta)=\frac{D^{4}\tilde{\theta}}{D\tilde{\theta}}-\frac{2D^{2}\tilde{\theta}D^{3}\tilde{\theta}}{(D\tilde{\theta})^{2}}\,.\label{eq:SS-N=00003D1}
\end{equation}
Expression \eqref{eq:barPsi} defines the coadjoint  action of the super-Virasoro group that reproduces its associativity. Indeed, by applying the supertransformation $\tilde{\theta}(\tau,\theta)$ with $\tau=\tau(\hat{\tau},\hat{\theta})$ and  $\theta=\theta(\hat{\tau},\hat{\theta})$ on $\tilde{\Psi}$, we recover the form of \eqref{eq:barPsi} since the super-Schwarzian \eqref{eq:SS-N=00003D1} satisfies the following cocycle condition
\begin{equation}
S(\tilde{\tau},\tilde{\theta};\hat{\tau}, \hat{\theta} )=(\hat{D}\theta)^{3}\, S(\tilde{\tau},\tilde{\theta} ; \tau,\theta)+S(\tau,\theta;\hat{\tau},\hat{\theta})\,.\label{eq:supercocy}
\end{equation}
Associativity of the group translates into closure of the algebra for linearized superdiffeomorphisms. Defining $\delta \Psi$ as the difference between of $\tilde{\Psi}$ and $\Psi$ at the same point $(\tau,\theta)$ and using  \eqref{infpa} on \eqref{eq:barPsi} yields 
\begin{equation}
\delta_{\chi}\Psi  =\frac{3}{2}D^{2}\chi\Psi+\frac{1}{2}D\chi D\Psi+\chi D^{2}\Psi+\frac{1}{2}D^{5}\chi\label{eq:dPsi}
\end{equation}
where $\chi(\tau,\theta)=\xi(\tau)-2\theta\epsilon(\tau)$, which is in agreement with the infinitesimal asymptotic symmetry transformations \eqref{eq:deltaL} and \eqref{eq:deltapsi}. 

The dilaton multiplet is characterized by the fields $y$ and $\rho$. In terms of superfields, this can be  organized into the following supervector
\begin{equation}
Y(\tau,\theta)\equiv y(\tau)-2\theta\rho(\tau)\,.\label{eq:def-Y}
\end{equation}
Specifically, $Y$ corresponds to an adjoint vector of the super-Virasoro group, then
\be
\tilde{Y}(\tilde{\tau},\tilde{\theta})=( D\tilde{\theta} )^{2} \, Y(\tau, \theta).
\ee
The infinitesimal action of the group on $Y$, $\delta Y= \tilde{Y}(\tau,\theta)-Y(\tau,\theta)$, is then given by 
\begin{equation}
\delta_{\chi}Y  =-D^{2}\chi Y+\frac{1}{2}D\chi D Y+\chi D^{2}Y\label{eq:dY}
\end{equation}
which reduces to the transformations \eqref{eq:delta y} and \eqref{eq:delta eps} obtained from the asymptotic analysis.

\section{Variational principle}\label{biendef}

A well-defined action principle for dilaton supergravity is given by the following functional
\begin{equation}
I=\frac{k}{2\pi}\intop\text{Str}\left[\textbf{X}\left(\extd\textbf{A}+\textbf{A}^{2}\right)\right]+I_{B}\,.\label{eq:I-wdef}
\end{equation}
In order to ensure that the variational principle
attains an extremum, the action \eqref{eq:BFaction} is supplemented with a surface
term $I_{B}$, that evaluated at $r\rightarrow\infty$ becomes
\begin{equation}
\delta I_{B}=-\frac{k}{2\pi}\ointop \extd\tau\,\text{Str}\left[\textbf{X\ensuremath{\delta}}\textbf{A}_{\tau}\right]\label{eq:dIB}
\end{equation}
where $\textbf{A}_{\tau}$ stands for the time component of the one-form $\textbf{A}$.
The variation of the action \eqref{eq:I-wdef} is then a bulk
term proportional to the equations of motion, that in consequence
vanishes on-shell. 

\subsection{Phase space of dilaton supergravity and integrability condition}\label{se:9.6}

The boundary term \eqref{eq:dIB} is a priori non-integrable and then one
must provide suitable integrability conditions at the boundary. Implementing the asymptotic conditions \eqref{eq:A}, \eqref{eq:X}
into the variation of $I_B$ \eqref{eq:dIB} yields
\begin{equation}
\delta I_{B}=\frac{k}{2\pi}\ointop \extd\tau\left(y\delta\mathcal{L}-2\rho\delta\psi\right)=\frac{k}{2\pi}\intop \extd\tau \extd\theta\, Y\delta\Psi\label{eq:deltaIB2}
\end{equation}
where $Y$ and $\Psi$ are given in \eqref{eq:def-Psi} and \eqref{eq:def-Y}, respectively. In the above  equation, we have introduced Berezin integrals defined by $\int \extd\theta=0$ and $\int \extd\theta\,\theta=1$.

In order to obtain the value of the boundary term $I_{B}$, one has to be able to integrate the expression \eqref{eq:deltaIB2}.  To accomplish this task, we need to single out a family of fields $(Y,\Psi)$ respecting our boundary conditions. A choice that is consistent with the asymptotic symmetry group corresponds to picking a set of fields $(y_0, \Psi_0)$ and act with a superreparametrization $\Theta$ to construct
\eq{
Y[\Theta; y_0]=\left(D{\Theta}\right)^{-2}y_0\qquad \qquad 
\Psi[\Theta;\Psi_0]=(D\Theta)^3\,\Psi_0+S\left(T,\Theta; \tau,\theta\right)
}{coadPsi}
where $T$ is the coordinate defined through the relation $DT = \Theta D\Theta$. Therefore, the phase space of $\cN=1$ dilaton supergravity is now defined  in terms of coadjoint orbits of the super--Virasoro group characterized by a representative $(y_0,\Psi_0)$ and a superdiffeomorphim $\Theta$. However, in order to maintain the phase space dimension, it is necessary to impose an additional restriction among these new variables. Here, we will define $\Theta$ as the transformation that brings the dilaton superfield $Y$ into its constant value $y_0$.  Furthermore, we demand as an integrability condition 
\be\label{integrability}
\delta y_0=0
\ee
meaning that $y_0$ is a constant without variation, which introduces a new coupling constant into the system. This condition has been used previously in the bosonic case \cite{Grumiller:2017qao,Gonzalez:2018enk}. In applications to the (super) SYK model in the low temperature regime, $1/y_0$ is the variance associated with the random quartic interaction \cite{Gonzalez:2018enk}.

Let us further describe phase space defined by \eqref{coadPsi} and \eqref{integrability}, by studying the equations of motion associated to it. Let us first note that 
\begin{equation}
y_0^2 D_{\Theta}\Psi_{0}=\cC + \theta \, y \, \delta_{(y,\rho)} \psi \label{eq:Integrability}
\end{equation}
with $D_{\Theta}=\d_{\Theta}+\Theta \d_{T}$ and $\cC$  is the asymptotic value of the Casimir function defined in \eqref{supercas}. On-shell, the latter equation can be easily solved using that $\cC$ is a constant of motion and that $\delta_{(y,\rho)} \psi$ vanishes, giving $\Psi_0\approx y_0^{-2} \Theta\, \cC+ \psi_0$ with $\psi_0$ a fermionic constant.\footnote{The expression $A\approx B$ means that $A$ and $B$ are equal provided the field equations are satisfied.} 

Summarizing, the superdiffeomorphism $\Theta$ is such that, when the field equations hold, it transforms  the conserved quantities of the system, $y_0$ and  $(\cC,\psi_0)$ into the fields $Y$ and $\Psi$, respectively. This is reminiscent of the Hamilton--Jacobi theory, where through a canonical transformation, the dynamics of the system is defined in terms of coordinates that are constants of motion. In the present case, the superreparametrization $\Theta$ plays the role of a canonical transformation. 

\subsection{Boundary term} \label{sec:boundaryterm}

Let us turn to the variation of $I_B$. Considering the family of fields \eqref{coadPsi} subject to the integrability condition \eqref{integrability}, the variation of the boundary action \eqref{eq:deltaIB2} becomes
\begin{equation}
\delta I_{B}=\frac{k y_0}{2\pi}\intop \extd\tau \extd\theta\left[\delta\left( D\Theta\, \Psi_{0}\right)+2D\delta\Theta \,\Psi_{0}+\left(D\Theta \right){}^{-2}\delta S\right]\,.\label{eq:dIB3}
\end{equation}
The third term in the above expression is dismissed since it is a total super derivative that vanishes due to the periodicity conditions on the superfield. For the second term, we use the on-shell condition $\Psi_0\approx \Theta y_0^{-2} \cC + \psi_0$, which transforms it into 
\begin{equation}
2 \cC \intop \extd\tau  \extd\theta \, \Theta \, D\delta \Theta= \cC \delta \intop  \extd \tau \extd\theta \, \Theta D \Theta  = \cC \delta \beta \, .\label{eq:dIB4}
\end{equation}
For a generic constant value of the Casimir $\cC$, the latter term vanishes by demanding that $\beta$ is a state--independent function. Another way of dealing with this term is considering configurations with $\cC =0$, and as will be seen in section \ref{sec:N1superconf}, this assumption turns out to be consistent with a boundary dynamics controlled by the action of the $\cN =1$ superparticle.

The first term in \eqref{eq:dIB3} defines a local expression for $I_B$. By
\emph{pulling out} the delta from it,  the boundary
term reads
\begin{equation}
I_{B}= \frac{k y_0 }{2\pi}\intop \extd\tau \extd\theta \, D\Theta \,  \Psi_0 \,.\label{eq:Ib}
\end{equation}
Having the boundary term $I_B$ to render the action well-defined allows us to compute the value of $I=I_{\rm bulk}+I_{B}$ for classical configurations. Once the equations of motion hold, $I_{\rm bulk}$ vanishes and the on-shell action value is fully determined by $I_{B}$, then
\be
\label{osv}
I_{\rm on-shell}=\frac{k \beta }{2\pi y_0}\cC
\ee
This expression is in agreement with the one obtained in the bosonic case \cite{Gonzalez:2018enk,Grumiller:2017qao}.

\section{Regularity conditions: Holonomy and super-Hill equation}\label{sec:Holonomy}

Euclidean black holes are smooth geometries as long as the time period is identified with the inverse of the Hawking temperature.  In a gauge formulation, this requires the holonomy $H_{C}[\mathbf{A}]$ to be trivial along a contractible cycle $C$. For $\mathfrak{osp}(2,1)$, the previous condition takes the form\footnote{See \cite{Henneaux:2015ywa} for an application of this condition in the context of hypersymmetric black holes in 3D.} 
\begin{equation}
H_{C}[\mathbf{A}]=\mathcal{P}e^{-\oint \mathbf{A}_{\mu}\extd x^\mu}=\Gamma_\pm=\left(\begin{array}{cc}
\pm\mathbb{I}_{2\times2} & 0_{2\times1}\\
0_{1\times2} & 1
\end{array}\right)\,.\label{eq:Holonomy1}
\end{equation}
Different signs in \eqref{eq:Holonomy1} are related to the possible spinorial representations; the center group element $\Gamma_{+}$ is consistent with fermions satisfying Ramond (periodic) boundary conditions, while $\Gamma_{-}$ is associated to fermions obeying Neveu-Schwarz (anti-periodic) boundary conditions.\footnote{Note that $\Gamma_{-}$ anticommutes with the fermionic generators
of $\mathfrak{osp}(2,1)$.}

In order to evaluate \eqref{eq:Holonomy1}, we consider the most general solution satisfying \eqref{eq:A} and \eqref{eq:X}. These are configurations of the form  $\mathbf{A}=b^{-1}(\extd+a)b$, where $b=\exp[\log(r)L_{0}]$ and $a=a_{\tau}\extd\tau$, with
\begin{equation}
a_{\tau}=L_{1}+\mathcal{L}(\tau)L_{-1}+\psi(\tau) G_{-\frac{1}{2}}\,.\label{eq:atau}
\end{equation}
Thus, in this case the holonomy condition becomes $H_{C}[a]=\Gamma_{\pm}$
and $C$ is chosen to be the thermal circle $0< \tau \leq  \beta$. Since we can always write the radial independent connection as $a=u\extd u^{-1}$,
the holonomy is reduced to $H_{C}[a]=u(\beta)u(0)^{-1}$. Regularity of the searched-for solutions requires to find a group element $u$ satisfying the boundary condition   $u(\beta)=\Gamma_{\pm}u(0)$. To do so, let us note that
$u$ can be written in terms of two independent three-dimensional row vectors, $\varphi_{(1)}$ and $\varphi_{(2)}$, as
\begin{equation}
u=\left(\begin{array}{c}
-\varphi_{(2)}\text{\ensuremath{'}}\\
\varphi_{(2)}\\
\varphi_{(1)}
\end{array}\right)
\end{equation}
where $\varphi_{(1)}$ and $\varphi_{(2)}$ are independent solutions to the super-Hill equation
\begin{equation}
\left(D^{3}+\Psi\right)\Phi=0\label{eq:superHill's}
\end{equation}
with $\Phi\equiv\varphi_{(2)}+\theta\varphi_{(1)}$ and $\Psi$ given in \eqref{eq:def-Psi}. 

When using the fact that \eqref{eq:superHill's} is covariant under the super-Virasoro group \cite{Bakas:1988mq,Mathieu:1990} and that $\Phi$ respects the boundary conditions for the group element, one can always perform a superdiffeomorphism and study solutions of \eqref{eq:superHill's} for a constant superfield $\Psi_\ast$.  Note that finding this superfield means that we are able to obtain solutions for a collection of densities $\Psi$ connected to $\Psi_\ast$ by a superdiffeomorphism. More precisely, $\Psi$ belongs to a super-Virasoro coadjoint orbit defined as quotient space between SDiff$(S^{1})$  modulo the transformations leaving $\Psi_\ast$ fixed. We characterize this orbit as follows. The solution of \eqref{eq:superHill's} for a constant $\Psi_\ast$ states that $\Phi$ is of the form $e^{\pm i\sqrt{\mathcal{L_\ast}}\tau}$. The boundary conditions on the group element $u(\beta)=\Gamma_{\pm}u(0)$ demand 
\begin{equation}
\mathcal{L}_\ast=\frac{4\pi^{2}n^{2}}{\beta^{2}}
\label{eq:angelinajolie}
\end{equation}
where $n$ is an arbitrary (half-)integer for (anti)periodic boundary conditions. In the Neveu-Schwarz sector the fermionic field $\psi_\ast$ has to vanish and the little group associated to $\Psi_\ast$ is a $2n$-cover of ${\rm OSp}(2,1)$. Thus, in the simplest case $n=\frac{1}{2}$, the superfield $\Psi$
stands for an element of the coadjoint orbit $\text{SDiff}(S^{1})/\text{OSp}(2,1)$.  

In the Ramond sector, there is not any condition on the fermionic field $\psi_\ast$. 
Generically, the isotropy group of $\Psi_\ast$ is U$(1)$. Then, $\Psi$ is a representative of the orbit  $\text{SDiff}(S^{1})/\text{U}(1)$. For the specific case $\cL_\ast=0$ the little group becomes a graded extension of U$(1)$, while for $\psi_\ast=0$, the little group corresponds to a $2n$-cover of ${\rm OSp}(2,1)$.

In order to make contact with the
Hawking temperature we will restrict, unless otherwise stated, to the Neveu-Schwarz sector and for definiteness we set $n=\frac{1}{2}$.

\section{Boundary dynamics}\label{bdf}

As concluded in section \ref{biendef}, the variational principle for
dilaton supergravity dictates that the bulk action has to be supplemented
by the boundary term \eqref{eq:Ib}. In this section it is shown that, for the considered class of configurations, this boundary term actually controls the boundary dynamics of the theory. For doing this, we will gradually reduce the action in two steps: 
\begin{enumerate}[$(i)$]
\item Evaluate the theory on the constraint surface, and make use of the asymptotic conditions  \eqref{AXExt1}.
\item Restrict the phase space to the one of regular solutions.
\end{enumerate}

\subsection[{\texorpdfstring{$\cN=1$}{N=1} super-Schwarzian action}]{\texorpdfstring{$\boldsymbol{\cN=1}$}{N=1} super-Schwarzian action} \label{sec:SchwN1}

In order to proceed with the proposed steps it is necessary to write the action in a Hamiltonian form, which can be done by splitting the field components in space and time. Thus, the variational principle \eqref{eq:I-wdef} becomes
\begin{equation}
I=-\frac{k}{2\pi}\int \extd\tau \extd r \,\text{Str}[\textbf{X}\textbf{A}'_r+\textbf{A}_{\tau}\cG]+\frac{k}{2\pi}\oint \extd\tau\, \text{Str}[xa_{\tau}]+I_{B}\label{eq:Iham1}
\end{equation}
where
\begin{equation}
\cG=\partial_{r}\textbf{X}+[\textbf{A}_r,\textbf{X}]\,.
\end{equation}
Note that the second term of \eqref{eq:Iham1} has been written in terms of the radial-independent fields $x$ and $a_\tau$, by virtue of the asymptotic conditions \eqref{AXExt1}.

Let us apply step  $(i)$. The variation of \eqref{eq:Iham1} with respect to the Lagrange multiplier $A_{\tau}$ implies the constraint $\cG=0$. Its solution is generically given by
\begin{equation}
\textbf{X}=G^{-1}x_0 G\label{SolX}
\end{equation}
with $x_0=x_0(\tau)$ and $G$ being a group element that satisfies $\textbf{A}_r=G^{-1}\partial_r G$.  Once the constraint is satisfied, the remaining bulk term appearing in \eqref{eq:Iham1} reduces to
\begin{equation}
-\frac{k}{2\pi}\oint \extd\tau \, \text{Str}[x_0 g'g^{-1}]\label{Boundary1}
\end{equation}
where we used that, according to \eqref{AXExt1}, the group element $G$ can be asymptotically factorized as $G=g(\tau)b(r)$. The group element $g$ can be determined from the fact that, for this class of configurations, the gauge connection is asymptotically flat, then $a_{\tau}=g^{-1}g'$. Additionally, one can also conclude that $x_0=g x g^{-1}$. By taking in consideration the above, \eqref{Boundary1} can be written as
\begin{equation}
-\frac{k}{2\pi}\oint \extd\tau\, \text{Str}[xa_{\tau}]
\end{equation} 
which is exactly cancelled with the second term in \eqref{eq:Iham1}. Hence, for the set of asymptotic conditions \eqref{AXExt1}, the value of the action on the constraint surface is solely determined by the boundary term $I_B$ given in \eqref{eq:Ib}. By virtue of \eqref{eq:barPsi}, the action \eqref{eq:Iham1} reduces to
\begin{equation}
I_{\text{red}}[\Theta; \Psi]=\frac{ky_0}{2\pi}\intop \extd\tau \extd\theta\left(D\Theta\right)^{-2}\left[\Psi-S\left(T,\Theta;\tau,\theta\right)\right]\,.\label{yquetanto}
\end{equation}
Here, the bulk inherited quantity $\Psi$ plays the role of a source for the superreparametrization field $\Theta$. At first glance, the reduced theory is invariant under the whole group of superdiffeomorphisms, whose infinitesimal transformation law for $\Psi$ is given in \eqref{eq:dPsi} and the one for $\Theta$ reads
\begin{equation}
\delta_{\chi}\Theta  =\chi \partial_{\tau} \Theta+\frac{1}{2} D\chi D\Theta \label{eq:dbarth}
\end{equation}
However, we still have to implement the step $(ii)$, which is performed as follows below. 

From the analysis made in section \ref{sec:Holonomy} [with $n=\tfrac12$ in \eqref{eq:angelinajolie}], regular solutions are given by elements whose representative
\begin{equation}
\Psi_\ast=\theta \frac{\pi^2}{\beta^2}
\end{equation}
belongs to a super--Virasoro coadjoint orbit.  In turn, the symmetry of \eqref{yquetanto} is broken to the set of configurations satisfying $\delta_{\chi} \Psi_\ast=0$, where  $\chi$ generates the $\mathfrak{osp}(2,1)$ algebra.

The boundary equations of motion that follow from \eqref{yquetanto} can be obtained by taking its variation with respect to the superreparametrization fields $\{T,\Theta\}$. These are given by $\delta_Y \Psi=0$, implying Casimir conservation $\partial_{\tau}\cC=0$, consistently with the variational principle.

 In order to make contact with a more familiar representation of the boundary action \eqref{yquetanto}, let us rewrite it in terms of the inverse superreparametrizations $\tau(T,\Theta)$, $\theta(T,\Theta)$. By means of the
identity,
\begin{equation}
S\left(T,\Theta;\tau,\theta\right)=-\left(D\Theta\right)^{3}S\left(\tau,\theta;T,\Theta \right)\,,
\end{equation}
we find that the boundary degrees of freedom are described by the \emph{$\cN=1$ super-Schwarzian theory}
\begin{equation}
I_{\text{red}}[\theta]=\frac{ky_0}{2\pi}\intop \extd T\extd\Theta \left[\frac{\pi^2}{\beta^2}\theta  \left(D_{\Theta} \theta\right)^{3}+S\left(\tau,\theta;T,\Theta \right)\right].\label{ssch}
\end{equation}
This action was shown to describe the Goldstone modes associated to the spontaneous superreparametrization symmetry breaking of the super-SYK model at the low temperature regime  \cite{Fu:2016vas}. In the second order formulation of minimal dilaton supergravity, \eqref{ssch} has been also obtained in \cite{Forste:2017apw} using a different regularization procedure than the one described here.

\subsection[\texorpdfstring{$\cN=1$}{N=1} superparticle ]{\texorpdfstring{$\boldsymbol{\cN=1}$}{N=1} superparticle } \label{sec:N1superconf}

In this subsection, we will show that for the class of configurations with a vanishing Casimir, the boundary dynamics of dilaton supergravity is rather governed by the $\cN=1$ superparticle model.
In order to obtain the corresponding action principle we recast the reparametrization field $\Theta$ in terms of a superfield $K=q-\theta \zeta$ of conformal dimension $-\frac{1}{2}$, such that $D\Theta=K^{-1}$, where $q$ and $\zeta$ are commuting and anticommuting functions, respectively. In fact, by virtue of \eqref{eq:dbarth}, the induced infinitesimal transformation law for $K$ is given by
\begin{equation}\label{supconf}
\delta_{\chi}K=-\frac{1}{2}\chi\text{\ensuremath{'}}K+\chi K\text{\ensuremath{'}}+\frac{1}{2}D\chi DK
\end{equation}
as expected for its conformal weight. Considering the above, the parametrization for the dynamical variables \eqref{coadPsi} becomes
\eq{
Y[K; y_0]=K^{2}y_0\qquad \qquad 
\Psi[K;\Psi_0]=K^{-3}\,\Psi_0 -K^{-1}DK'\,.
}{conformalPsi}
By replacing this change of variables in \eqref{eq:deltaIB2} and taking under consideration the integrability condition \eqref{integrability}, the variation of the boundary term is now given by
\begin{equation}
\delta I_{B}=\frac{k y_0}{2\pi}\intop \extd\tau \extd\theta\left[\delta \left( K^{-1}\, \Psi_{0}\right)-2\Psi_{0}\,\delta[\log(K)]-K^{2}\delta(K^{-1}DK')\right]\,.\label{eq:dIBconf}
\end{equation}
The third term of \eqref{eq:dIBconf} is a total derivative and thus is dismissed. The second term cannot be integrated for arbitrary $\Psi_0$. When restricting to configurations satisfying $\Psi_0=0$ we can get rid of this term. This is, in fact, the only possible choice consistent with supersymmetry (recall that $\Psi_0$ is an odd superfield, while $K$ is even). In that case, only the first term contributes to $I_B$, obtaining that its local expression is given by
\begin{equation}
I_{B}= \frac{k y_0 }{2\pi}\intop \extd\tau \extd\theta \, K^{-1}\,  \Psi_0 \,.\label{eq:Ibconf}
\end{equation}
The on-shell value of the bulk action for classical configurations is only given by the boundary term, which under the assumption $\Psi_0=0$ identically vanishes
\be
\label{osvconf}
I_{\rm on-shell}=0\,.
\ee

The residual dynamics is obtained by using the steps $(i)$ and $(ii)$ described at the beginning of this section. The action on the constraint surface gives
\begin{equation} \label{N1-sc-action}
I_{\text{red}}[K;\Psi]=\frac{ky_{0}}{2\pi}\int \extd\tau \extd\theta\left(K^{2}\Psi_\ast-K\text{\ensuremath{'}}DK\right)
\end{equation}
which, at first sight, seems to be the superfield form of  the $\cN=1$ superconformal mechanics action without a self- interacting potential \cite{Fubini:1984hf}
\begin{equation}\label{boundconf1}
I_{\text{red}}[q,\zeta;\mathcal{L}_\ast,\psi_\ast]=\frac{ky_{0}}{2\pi}\oint \extd\tau\left[\cL_\ast q^{2}+2\psi_\ast q\zeta-(q')^{2}+\zeta\zeta'\right]\,.
\end{equation}
The action \eqref{N1-sc-action} is invariant under SDiff$(S^{1})$. This is because of the presence of the external source $\Psi_\ast$, which possesses conformal weight $\frac{3}{2}$. However, regular bulk configurations must respect the integrability condition $\Psi_0=0$. This restricts us to the Ramond sector of the holonomy operator \eqref{eq:Holonomy1}, whose little group is graded U$(1)$ with representatives $\cL_\ast=\psi_\ast=0$. Hence, the boundary model effectively describes the free $\cN=1$ superparticle .

Let us study the dynamics of the model. Taking the variation of \eqref{N1-sc-action} with respect to $K$ yields the equations of motion
\begin{equation}
K\Psi_\ast+DK\text{\ensuremath{'}}=0
\end{equation}
which, by virtue of \eqref{conformalPsi}, directly imply $\Psi_0=0$ or similarly $\cC=0$. This fact ensures that the parametrization of the boundary phase space in terms of the superdensity $K$ \eqref{conformalPsi} is consistent with the variational principle.

Finally, we mention that $\cC =0$ coincides with the existence condition of BPS-like states in the context of graded Poisson sigma models \cite{Bergamin:2003mh}. Therefore, bulk configurations with unbroken supersymmetries necessarily reduce their dynamics to a free superparticle model.

\section{Extended case}\label{extended}

\subsection{Extended dilaton supergravity}
Dilaton supergravity in 2D can be easily generalized in order to accommodate $N$ gravitini $\psi_A$ corresponding to Majorana spinors. This can be achieved by enlarging the gauge group of the BF action \eqref{eq:BFaction} to OSp$(2,N)$, where it is necessary to introduce additional SO$(N)$ gauge fields $Y^{AB}$ and $C^{AB}$.  In this case, $\textbf{X}$ and $\textbf{A}$, given by
\begin{align}
\textbf{X} & =X^{I}J_{I}+\lambda_{A}^{\alpha}Q_{\alpha}^{A}+\frac{1}{2}Y^{AB}T_{AB}\label{eq:XN}\\
\textbf{A} & =A^{I}J_{I}+\psi_{A}^{\alpha}Q_{\alpha}^{A}+\frac{1}{2}C^{AB}T_{AB}\label{eq:AN}
\end{align}
are evaluated on the $\mathfrak{osp}\left(2,N\right)$
algebra:
\begin{gather}
\begin{align}
[J_{I},J_{J}]&=\epsilon_{IJK}J^{K} & [J_{I},Q_{\alpha}^{A}]&=\frac{1}{2}(\Gamma_{I})_{\,\,\,\alpha}^{\beta}Q_{\beta}^{A}\\
[Q_{\alpha}^{A},T^{BC}]&=2\delta^{A[B}Q_{\alpha}^{C]} & [T^{AB},T^{CD}]&=4\delta^{[A|[C}T^{D]|B]}
\end{align}\\
\{Q_{\alpha}^{A},Q_{\beta}^{B}\}=-\frac{1}{2}\delta^{AB}\left(C\Gamma^{I}\right)_{\alpha\beta}J_{I} -\frac{1}{4}C_{\alpha\beta}T^{AB}\,.
\end{gather}
 The capital latin letters $A,B,C,D,\dots$, that label the generators $T_{AB}=-T_{BA}$ of the SO$(N)$ subgroup, run from $1$ to $N$.

We choose the supertrace whose nonvanishing components are given by (see appendix \ref{sec:A1})
\begin{equation}
\text{Str}[J_{I}J_{J}]=\frac{1}{2}\eta_{IJ}\quad\qquad\text{Str}[Q_{\alpha}^{A}Q_{\beta}^{B}]=\frac{1}{2}C_{\alpha\beta}\delta^{AB}\quad\qquad\text{Str}[T^{AB}T^{CD}]=4\delta^{A[C}\delta^{D]B}\,.\label{eq:StrN-1}
\end{equation}

Taking into consideration the above definitions, one obtains that the field strength associated with the gauge field $\textbf{A}$ reads
\begin{equation}
\textbf{F} =\big(R^{I}-\frac{1}{4}\bar{\psi}_{A}\Gamma^{I}\psi_{A}\big)J_{I}+\nabla\psi_{A}^{\alpha}Q_{\alpha}^{A}+\frac{1}{2}F^{AB}T_{AB}\,.\label{eq:FN}
\end{equation}
Here, the curvature $R^{I}$ is given in the first equation of \eqref{Rypsi}, while the covariant derivative acting on spinors and the
field strength associated to $C$ are given by
\begin{align}
\nabla\psi_{A}^{\alpha} & =d\psi_{A}^{\alpha}+\frac{1}{2}A^{I}(\Gamma_{I})_{\,\,\,\beta}^{\alpha}\psi_{A}^{\beta}+C_{AB}\psi_{B}^{\alpha}\\
F^{AB} & =dC^{AB}+C^{AC}C^{CB}-\frac{1}{2}\bar{\psi}^{[A}\psi^{B]}
\end{align}
respectively. By replacing \eqref{eq:XN}, \eqref{eq:AN} into \eqref{eq:BFaction}
and then using the supertrace \eqref{eq:StrN-1}, the action for the extended supergravity theory becomes
\begin{equation}
I=\frac{k}{2\pi}\int X^{I}\big[\big(R_{I}-\frac{1}{4}\bar{\psi}_{A}\Gamma_{I}\psi_{A}\big)+\bar{\lambda}_{A}\nabla\psi_{A}+Y_{AB}F^{AB}\big]\,.\label{eq:IN}
\end{equation}
As in the $\mathcal{N}=1$ case, the field equations \eqref{eq:BFeqns} impose that the dilaton $\textbf{X}$ must be the stabilizer of the flat gauge connection $\textbf{A}$.

Action \eqref{eq:IN} is invariant under the following local extended supersymmetry transformations
\begin{align}
\delta A^{I}&=-\frac{1}{2}\bar{\psi}_{A}\Gamma^{I}\epsilon_{A} & \delta X^{I}&=-\frac{1}{2}\bar{\lambda}_{A}\Gamma^{I}\epsilon_{A}\label{eq:d1-N}\\
\delta C^{AB}&=-\frac{1}{2}\bar{\psi}^{[A}\epsilon^{B]} & \delta Y^{AB}&=-\frac{1}{2}\bar{\lambda}^{[A}\epsilon^{B]}\label{eq:d2-N}\\
\delta\psi_{A}&=\nabla\epsilon_{A} & \delta\lambda_{A}&=\frac{1}{2}X^{I}\Gamma_{I}\epsilon_{A}+Y_{AB}\epsilon{}_{B}\label{eq:d3-N}
\end{align}
which can be obtained from \eqref{eq:Gaugetrafos} by considering the gauge parameter $\eta=\epsilon_{A}^{\alpha}Q_{\alpha}^{A}$.

\subsection{Asymptotic conditions}\label{sec:Asymptotic conditions 2}

The objective of this subsection is to study asymptotic configurations that are left invariant under the group of extended superdiffeomorphisms on the circle.
Such group comprises diffeomorphisms $\xi(\tau)$, local supersymmetry transformations $\eps^A(\tau)$ and internal SO$(N)$ gauge symmetries $\mu^{AB}(\tau)$. The analysis of the asymptotic symmetries is performed by imposing the condition \eqref{eq:Gaugetrafos} modulo trivial gauge transformations of the form \eqref{trivial}.

The configurations of interest can be expressed as in \eqref{AXExt1} where the gauge field $a$ and the radial independent dilaton $x$ are given by 
\begin{align}
\label{AXExt2}
a&=\big[L_{1}+\big(\mathcal{L}+\frac{1}{2}\mathcal{P}^{AB}\mathcal{P}^{AB}\big)L_{-1}+\mathcal{\psi}^{A}G_{-\frac{1}{2}}^{A}+\frac{1}{2}\mathcal{P}^{AB}T^{AB}\big]\, \extd\tau \\ \nonumber 
x&=y L_{1}-y'L_{0}+\big[\frac{1}{2}y\text{\ensuremath{''}}+y\big(\mathcal{L}+\frac{1}{2} \cP^{AB} \cP^{AB} \big)+\psi^{A}\rho^{A}\big]L_{-1} \\
&\quad +\frac{1}{2}\left(\lambda^{AB}+y\mathcal{P}^{AB} \right)T^{AB}+\rho^{A}G_{\frac{1}{2}}^{A}-\left(\rho^{A}\text{\ensuremath{'}}-y\mathcal{\psi}^{A}+\mathcal{P}^{AB}\rho^{B}\right)G_{-\frac{1}{2}}^{A}\,. \label{AXExt3}
\end{align}
This set of fall--off conditions is a consistent extension of the minimal supersymmetric case \eqref{eq:A} and \eqref{eq:X}. Here, we have considered the following basis for the $\mathfrak{osp}(2,N)$ algebra (see appendix \ref{sec:A1})
\begin{gather}
\label{AXExt4}
\begin{align}
\left[L_{m},L_{n}\right]&=\left(m-n\right)L_{m+n} & \left[L_{m},G_{p}^{A}\right]&=\left(\frac{m}{2}-p\right)G_{m+p}^{A}\\
[G_{p}^{A},T^{BC}]&=2\delta^{A[B}G_{p}^{C]} & [T^{AB},T^{CD}]&=4\delta^{[A|[C}T^{D]|B]}
\end{align}\\
\{G_{p}^{A},G_{q}^{B}\}=-2L_{p+q}\delta^{AB}-(p-q)T^{AB}\,.
\end{gather}
with $m,n=\pm1,0$ and $p,q=\pm\frac{1}{2}$. 
As explained in section \ref{sec:Asymptotic conditions}, the Lie-algebra-parameter
\be
\Lambda=\textbf{X}[\xi,\epsilon_{A},\mu_{AB}]\label{AXExt5}
\ee
singles out the gauge transformations that keep invariant the asymptotic behavior \eqref{AXExt2} and \eqref{AXExt3} as long as the transformation laws of the fields take the following form
\begin{align}
\delta_{(\xi,\epsilon,\mu)}\mathcal{L} & = 2\mathcal{L}\xi\text{\ensuremath{'}}+\mathcal{L}\text{\ensuremath{'}}\xi+\frac{1}{2}\xi\text{\ensuremath{'''}}+3\psi^{A}\epsilon^{A}\text{\ensuremath{'}}+\psi^{A}\text{\ensuremath{'}}\epsilon^{A}-\mathcal{P}^{AB}\mu^{AB}\text{\ensuremath{'}} \label{dL} \\ 
\delta_{(\xi,\epsilon,\mu)}\psi^{A} & =  \frac{3}{2}\psi^{A}\xi\text{\ensuremath{'}}+\psi^{A}\text{\ensuremath{'}}\xi-\mu^{AB}\psi^{B}-\mathcal{L}\epsilon^{A}-\eps^{A}\text{\ensuremath{''}}-\mathcal{P}^{AB}\text{\ensuremath{'}}\epsilon^{B} -2\mathcal{P}^{AB}\epsilon^{B}\text{\ensuremath{'}}\nonumber\\ \label{dp}
&\quad -\big(\cP^{AB}\cP^{BE}+\frac{1}{2}\mathcal{P}^{BC}\mathcal{P}^{BC}\delta^{AE}\big)\eps^{E} \\
\delta_{(\xi,\epsilon,\mu)}\mathcal{P}^{AB} & =  \mathcal{P}^{AB}\xi\text{\ensuremath{'}}+\mathcal{P}^{AB}\text{\ensuremath{'}}\xi+2\mathcal{P}^{[A|C}\mu^{C|B]}+\mu^{AB}\text{\ensuremath{'}}+2\psi^{[A}\epsilon^{B]}\label{dP}\\
\delta_{(\xi,\eps,\mu)}y & =  y' \xi - y \xi'  -2 \rho_A \epsilon_A   \label{eq:Y1}\\
\delta_{(\xi,\eps,\mu)}\rho_A &= \rho_A' \xi - \frac{1}{2} \rho_A  \xi' +\frac{1}{2}y' \epsilon_A-y\epsilon'_A+ \rho_{B}\mu_{BA}+\lambda_{AB}\eps_B \label{eq:Y2}\\
\delta_{(\xi,\eps,\mu)} \lambda_{AB} &=\xi \lambda'_{AB}-y\mu_{AB}'  + 2\lambda_{C[A}\mu_{B]C} + 2 \rho_{[A} \epsilon_{B]}'-2\rho_{[A}' \eps_{B]}+2\rho_{C} \cP_{C[A}\eps_{B]}  \label{eq:Y3} \nonumber\\
&\quad -2\eps_{C}  \cP_{C[A}\rho_{B]}+2\rho_{C}\epsilon_{C}\cP_{AB} \, .
\end{align}
This set of infinitesimal transformations defines the action of the $SO(N)$--extended super--Virasoro algebra \cite{Bershadsky:1986ms,Knizhnik:1986wc} on the dynamical fields. In the generic case, the transformations are quadratic in the affine
$\mathfrak{so}(N)$ currents $\mathcal{P}^{AB}$. However, for
$\mathcal{N}=2$ the non-linear terms appearing in $\delta \psi^{A}$ and $\delta \lambda_{AB}$
cancel out each other, leaving the transformation laws linear.  

As was shown in section \ref{bdf}, the boundary dynamics of the $\cN=1$ case is controlled by a finite superdiffeomorphism $\Theta$. This is a property associated to the asymptotic symmetry group. Due to the non-linearities present in \eqref{dp}, the corresponding group structure is not completely understood for generic $N$. Since only the $SO(2)$--extended super-Virasoro algebra reproduces linear transformations in the fields, in the remainder of this section we will restrict ourselves to that case only.

Even though we do not have a representation for finite transformations with $N>2$, one can use the relation between dilaton gravity and the action of the particle on a group manifold \cite{Brigante:2002rv,Mertens:2018fds,Gonzalez:2018enk} to find suitable boundary theories for the breaking of the $SO(N)$--extended superreparametrization groups. This connection will be studied in section \ref{ultimasec}.

\subsection[\texorpdfstring{$\cN=2$}{N=2} superspace formulation]{\texorpdfstring{$\boldsymbol{\cN=2}$}{N=2} superspace formulation}

Hereafter, we will make use of the extended superspace formalism to represent the field content and the action of the asymptotic symmetry group. In accordance with \cite{Schoutens:1988ig}, we will denote the supercoordinates as the set $z=(\tau, \theta_1,\theta_2)$. The corresponding covariant derivative $D_{A}= \theta_{A} \d_{\tau}+ \d_{A}$ satisfies
\be\label{eq:AXExt7}
\{D_A, D_B \}= 2 \delta_{AB} \d_\tau\,.
\ee
The requirement that $D_A$ transforms covariantly, i.e., $D_{A}=\left(D_{A} \tilde{\theta}_B \right)\tilde{D}_{B}$  leads to the following  constraints 
\be
D_{A}\tilde{\tau}= \tilde{\theta}_{B} D_{A} \tilde{\theta}_B \qquad \qquad 
(D_A \tilde{\theta}_C) (D_B\tilde{\theta}_C)= \delta_{AB}\left( \d_\tau\tilde{\tau}+\tilde{\theta}_C  \d_\tau\tilde{\theta}_C\right)\,. \label{IdN1}
\ee
An infinitesimal superreparametrization can be written as $\delta \tau= -\chi +\frac{1}{2} \theta_A D_A \chi$ and  $\delta \theta_A= -\frac{1}{2} D_A \chi$,  with $\chi=\xi- 2\theta_{B}\epsilon_B + \theta_{A} \theta_{B} \mu_{AB}$. It is then found  that $\tilde{\theta}_A$ transforms as
\be
\delta \tilde{\theta}_A= \chi \d_\tau {\tilde{\theta}_{A}} +\frac{1}{2} \left(D_B \chi \right) D_B\tilde{\theta}_A\,. \label{lamagia}
\ee
The field content can be arranged in the following superfields
\begin{align}
\Psi&=\theta_1\theta_2 \cL + \epsilon_{A B} \big( \psi_{A} \theta_{B} -\frac{1}{2}\cP_{A B} \big)\label{eq:PsiN2}\\
Y&=y- 2\theta_{B}\rho_B + \theta_{A} \theta_{B} \lambda_{AB}\label{eq:YN2}
\end{align}
so that it reproduces the action of the asymptotic symmetry group. In the case of $\Psi$, it can be seen that its finite transformation corresponds to the coadjoint action of the $\cN=2$ super-Virasoro group \cite{Bakas:1988mq,Delius:1990pt,Mathieu:1990,YangKQ} on a quasi-primary field of conformal weight one. The coadjoint action is dictated by
\be
\tilde{\Psi}(\tilde{z})=\Omega^{-1}[\Psi(z)-S^{(2)}(\tilde{z};z)]  \qquad \qquad S^{(2)}(\tilde{z};z)= \frac{1}{2}\left(\frac{D_1 D_2 \Omega }{ \Omega }-\frac{3}{2} \frac{D_1\Omega D_2\Omega }{\Omega^2} \right) \label{coad2}
\ee
where $\Omega$ stands for the determinant of $D_A \tilde{\theta}_B$ and $S^{(2)}$ is the generalization of the Schwarzian derivative. Infinitesimally, \eqref{coad2} reproduces the transformations of the fields \eqref{dL}, \eqref{dp}, \eqref{dP} for the $N=2$ case by means of \eqref{lamagia}.  Furthermore, under the same group action, $Y$ transforms as
\be
\tilde{Y}(\tilde{z})=\Omega \, Y(z)\,. \label{YadN2}
\ee
from which it is recognized that $Y$ is a vector superfield.  Noteworthy, by virtue of \eqref{lamagia}, the infinitesimal version of  \eqref{YadN2} reproduces \eqref{eq:Y1}, \eqref{eq:Y2}, \eqref{eq:Y3}. 

\subsection[Action principle for \texorpdfstring{$\cN=2$}{N=2} dilaton supergravity]{Action principle for \texorpdfstring{$\boldsymbol{\cN=2}$}{N=2} dilaton supergravity}
It is necessary to  provide a well--defined variational principle for \eqref{eq:IN} associated to the set of extended asymptotic conditions \eqref{AXExt1}. As before, the enhanced action, given by the bulk term \eqref{eq:IN} plus a boundary term $I_B$, has an extremum as long as
\begin{equation}
\delta I_{B}=-\frac{k}{2\pi}\ointop \extd\tau\,\text{Str}\left[\textbf{X\ensuremath{\delta}}\textbf{A}_{\tau}\right]\,=\frac{k}{2\pi}\ointop \extd\tau\, \left(y \delta \cL -\lambda^{AB}\delta \cP^{AB} - 2\rho^{A} \delta \psi^{A} \right)\, .\label{eq:dIB22}
\end{equation}
By using the superfields defined in \eqref{eq:PsiN2} and \eqref{eq:YN2} we can can express the above integral as
\begin{equation}
\delta I_{B}=\frac{k}{2\pi}\intop \extd\tau\, \extd\theta_1 \,  \extd\theta_2 \, Y \delta \Psi \label{megustaria}
\end{equation}
where  Berezin integration  is defined by $\int  \extd\theta_1  \, \extd\theta_2 \, \theta_1 \, \theta_2=1$ and  $\int \extd\theta_1 \, \extd\theta_2=0$. 

\subsubsection{Phase space}
Analogously to $\cN=1$, we  parametrize the phase space by means of $\cN=2$ super-Virasoro coadjoint orbits. Then, the superfields $Y$ and $\Psi$ can be written as
\eq{
Y=\Omega_{\Theta}^{-1} y_0 \qquad\qquad \Psi=\Omega_{\Theta} \Psi_0 + S(T, \Theta_A; z)
}{coadN}
where $\Omega_{\Theta}=\det(D_{A} \Theta_B)$ and $T$ is the coordinate associated to the constraint $D_CT = \Theta_A D_C\Theta_A$. Here, we choose  $\Theta_A$ to be the reparametrization that brings $Y$ into a constant $y_0$. In what follows we will assume that $y_0$ is a fixed constant without variation, i.e., $\delta y_0=0$. 

Before analyzing the boundary term that makes the variational principle well-defined, let us discuss the relation between $\Psi_0$ and the conserved charges when the field equations are satisfied. If one takes two successive supercovariant derivates to $\Psi_0$ with respect to $\Theta_2$ and $\Theta_1$, we find that 
\be
\label{ultima}
 y_{0}^{2}D_{\Theta_1} D_{\Theta_2} \Psi_0 = -\cC - \theta_A ( y \delta_{(y,\rho,\lambda)} \psi_A+ 2 \rho_B \delta_{(y,\rho,\lambda)} \cP_{A B} ) - \frac{1}{2}\theta_A\theta_B(y \delta_{(y,\rho,\lambda)} \cP_{AB})'
\ee
where $\cC$ is the asymptotic value of the Casimir, that by virtue of its definition in \eqref{cas} is given by
\beq \label{CasimirN2}
\cC=\mathcal{L}y^{2}+3y\psi_A\rho_A+(2\rho_{A}\rho_{B}-y\lambda_{AB})\mathcal{P}_{AB}+\frac{1}{2}yy''-\frac{1}{4}\left(y'\right)^{2}+2\rho_A \rho'_A-\frac{1}{2}\lambda_{AB}\lambda_{AB}\,.
\eeq
On-shell, equation \eqref{ultima} can be solved to obtain
\be
\Psi_0\approx y_0^{-2}\Theta_1 \Theta_2 \cC + \epsilon_{AB}(\psi^{(0)}_A \Theta_{B}- \frac{1}{2} \cP^{(0)}_{AB})\label{integraN}
\ee
where $\psi^{(0)}_A$ and $\cP^{(0)}_{AB}$ are two extra possible on-shell constants. We can again conclude that $\Theta_A$ brings $\Psi$ to the frame where its components are defined in terms of the constants of motion $(\cC,\psi^{(0)}_A,\cP^{(0)}_{AB})$.

\subsubsection{Boundary term}
\label{telodijeviejito}

Let us go back to \eqref{megustaria} using the reduced phase space defined by \eqref{coadN}. Plugging in these expressions, it is found that
\be
\delta I_{B}=\frac{k y_0}{2\pi}\intop \extd z \left[\delta \Psi_{0}+ \Omega_{\Theta}^{-1}\delta \Omega_{\Theta} \Psi_{0}+ \Omega_{\Theta}^{-1} \delta S^{(2)}\right]\,.\label{IBN}
\ee
Here, the last term is a total derivative that vanishes due to the periodicity of the fields. In order to provide a local expression for $I_B$ we need to deal with the integrability problem present in the second term of \eqref{IBN}. Imposing the on-shell condition \eqref{integraN}, this term becomes 
\be
\label{zeroN2}
\intop \extd z  \Omega_{\Theta}^{-1}\delta \Omega_{\Theta}  \Psi_{0} \approx y_0^{-2} \cC \intop  \extd z \, \Theta_{1} \Theta_{2}  \,  D_{\Theta_A} \delta \Theta_A \,.
\ee
We must express the latter integral in terms of $T$ and $\Theta_A$. Since the Berezinian between $z$ and $(T,\Theta_A)$  is the identity, the right-hand side of \eqref{zeroN2}  is zero since $\beta$ is a state independent function. Bear in mind that this derivation is only  valid for the case $\cC \neq 0$. As it will be explained in section \ref{N2sqm}, the case $\cC=0$ needs an special treatment and in fact the term in \eqref{zeroN2} will yield a new contribution to $I_{B}$. Indeed, it will be proved that this enhancement leads to the action of  $\cN=2$ superconformal mechanics.

The remaining term in \eqref{IBN} defines the value of the boundary term
\be
I_B=\frac{k y_0}{2\pi} \intop \extd \tau \extd\theta_1 \extd\theta_2 \, \Psi_0\,.
\label{magic}
\ee
The value of the on-shell action is
again given by the boundary term $I_B$. In terms of the Casimir function this value takes the same the form as in \eqref{osv}.

\subsection{Regularity conditions: extended super-Hill equation}\label{HoloN2}

The asymptotic symmetry breaking exhibited in the case with $\cN=1$ replicates in the model with extended supersymmetries. For seeing this, let us proceed to find the class of configurations satisfying trivial holonomy around the thermal circle $C$. The generalization of the center group element to OSp$(2,N)$ is given by
\begin{equation}
\Gamma_{\pm}=\left(\begin{array}{cc}
\pm\mathbb{I}_{2\times2} & 0_{2\times N}\\
0_{N\times2} & \mathbb{I}_{N \times N}
\end{array}\right)\,.\label{eq:HolonomyN}
\end{equation}
Thus, from $a=udu^{-1}$ one translates the trivial holonomy condition $H_{C}[a]=\Gamma_{\pm}$  to the boundary condition for the group element $u(\beta)=\Gamma_{\pm}u(0)$.

In order to determine $u$ for the case at hand, we have to consider the expression of the gauge connection $a$ \eqref{AXExt2} and the matrix representation of the OSp$(2,2)$ generators (see appendix \ref{sec:A1}). It is then found that, in terms of $3$ independent $4-$dimensional row vectors $\varphi_{(1)}^A$ and $\varphi_{(2)}$, the group element reads
\begin{equation}
u=\left(\begin{array}{c}
-\varphi_{(2)}\text{\ensuremath{'}}\\
\varphi_{(2)}\\
\varphi_{(1)A}
\end{array}\right)
\end{equation}
where $\varphi_{(1)A}$ and $\varphi_{(2)}$ are solutions to the $\mathcal{N}=2$ super-Hill
equation
\begin{equation}
\left(D_{1}D_{2}+\Psi\right)\Phi=0\,.\label{eq:superHillN2}
\end{equation}
Here $\Phi\equiv\varphi_{(2)}+\theta_{A}\varphi_{(1)A}-\theta_{1}\theta_{2}\cP \varphi_{(2)}$, with $\cP=\frac{1}{2}\epsilon_{AB}\cP_{AB}$ and $\Psi$ given
in \eqref{eq:PsiN2}. Thus, the components of $\Phi$ must satisfy the following equations 
\begin{align}
\varphi_{(2)}''+(\cL +\cP ^2)\varphi_{(2)}-\psi_A \varphi_{(1)A}&= 0\,\label{Hill1}\\ 
\varphi_{(1)A}'+\cP \epsilon_{AB}\varphi_{(1)B}+\psi_A \varphi_{(2)}&= 0\,.\label{Hill2}
\end{align}
For solving \eqref{eq:superHillN2}, we will use the fact that the extended super-Hill equation is covariant under the group of extended superreparametrizations \cite{Bakas:1988mq,Mathieu:1990}, one is then allowed to take the system to the rest frame, where the supercurrent $\Psi=\Psi_\ast$ is constant. In the Neveu-Schwarz sector, $u(\beta)=\Gamma_{-}u(0)$, it is found that $\psi^{A}_\ast$ must vanish and thus the above equations decouple. It follows that
\begin{equation}
\label{eq:22}
\cL_\ast= \frac{4\pi^2}{\beta^2}\left(n^2-m^2\right)\qquad\qquad \cP_\ast=\frac{2\pi m}{\beta}
\end{equation}   
where $n$ and $m$ are restricted to be half-integers and integers, respectively. In the case $m=0$, the little group of $\Psi_\ast$ is given by a $2n$-cover of OSp$(2,2)$. In consequence, for $n=\frac{1}{2}$, $\Psi$ corresponds to an element of the coadjoint orbit with OSp$(2,2)$ as little group \cite{Bakas:1988mq,YangKQ}. 

In the Ramond sector, $\Gamma_{+}=\mathbb{I}_{4\times 4}$, the solutions to the equations \eqref{Hill1} and \eqref{Hill2} are of the form $\exp (i\lambda_{\pm} \tau)$,  with $\lambda_+=\frac{2\pi n}{\beta}$ and $\lambda_-=\frac{2\pi m}{\beta}$ with $n$  and $m$ arbitrary integers.  Making use of the \textit{body and soul} decomposition $\lambda_{\pm}=\lambda^{\pm}_{\rm b}+\lambda^{\pm}_{\rm s}\psi^{1}_\ast\psi^{2}_\ast$ in \eqref{Hill1} and \eqref{Hill2} we find that $\lambda^{\pm}_{\rm b}$ and $\lambda^{\pm}_{\rm s}$ satisfy
\begin{equation}
\lambda^{+}_{\rm b}=\pm\sqrt{\cL_\ast+\cP^2_\ast}\,,\quad \lambda^{-}_{\rm b} =\pm\cP_\ast \,, \quad \left[\cP_\ast-(2(\lambda^{\pm}_{\rm b})^2-\cL_\ast-2\cP^2_\ast)\lambda^{\pm}_{\rm b}\lambda^{\pm}_{\rm s}\right]\psi^{1}_\ast\psi^{2}_\ast=0\,.
\end{equation}
The cases $\cL_\ast=0$ and $\cL_\ast=-\cP_\ast^2$ deserve special attention. The case $\cL_\ast=0$  necessarily implies that $\cP_\ast=0$, without any restrictions on the fermions $\psi^A_\ast$. For $\psi^{1}_\ast=0$ (or $\psi^{2}_\ast=0$), the corresponding isotropy group is graded U(1).  If  $\cL_\ast=-\cP_\ast^2$ (with $\cP_\ast\neq 0$), the boundary conditions  imply that $\psi_\ast^1\psi_\ast^2=0$. This case can be represented by \eqref{eq:22} setting $n=0$ (and $m$ integer). Coadjoint orbits in this case are again characterized by graded U$(1)$.

For the generic case, $\cL_\ast\neq0$ and $\cL_\ast\neq -\cP_\ast^2$, one finds that 
\begin{equation}
\lambda_+=\pm\bigg(\sqrt{\cL_\ast+\cP^2_\ast}+\frac{\cP_\ast \psi^{1}_\ast\psi^{2}_\ast}{\cL_\ast\sqrt{\cL_\ast+\cP^2_\ast}}\bigg) \qquad \qquad \lambda_- =\pm\bigg(\cP_\ast -\frac{\psi^1_\ast \psi^2_\ast}{\cL_\ast}\bigg)\,.
\end{equation}
If $\psi_\ast^1\psi_\ast^2=0$, we recover \eqref{eq:22} (with $n$  integer). There are no additional conditions on $\psi^A_\ast$, therefore the superfield $\Psi_\ast$ is preserved only by U$(1)$ transformations. In the case of $\psi^{A}_\ast=0 $ and $\cP_\ast=0$, the isotropy group of $\Psi_\ast$  is given by a $2n$-cover of OSp(2,2).

\subsection[\texorpdfstring{$\cN=2$}{N=2} super-Schwarzian theory]{\texorpdfstring{$\boldsymbol{\cN=2}$}{N=2} super-Schwarzian theory}
\label{N2sqm}

As in the $\cN=1$ case, shown in section \ref{bdf}, the variational principle for the $\cN=2$ case evaluated on the constraint surface is also determined by the boundary term $I_{B}$, given by \eqref{magic}. After using \eqref{coad2} to express $\Psi_0$ in terms of $\Psi$, we find
\begin{equation}\label{onshell2}
I_{\rm red}=\frac{k y_0}{2\pi}\int \extd\tau \extd\theta_{1}\extd\theta_{2}  \; \Omega_{\Theta}^{-1}\left[\Psi-S^{(2)}(T, \Theta_A; z)\right].
\end{equation}
The equations of motion are obtained varying with respect to $\{T,\Theta_A\}$, conditions that imply the conservation of the boundary Casimir, $\d_{\tau}\cC=0$, being in complete agreement with the bulk dynamics.
Additionally, the action \eqref{onshell2} is invariant under the following transformations
\begin{align}
\delta \Psi&=\partial_{\tau} \chi\Psi+ \chi \partial_{\tau}\Psi+\frac{1}{2}D_{A}\chi D_{A}\Psi+\frac{1}{2}D_{1}D_{2}\partial_{\tau}\chi\label{eq:dpsi2}\\ 
\delta\Theta_A & =\chi \partial_{\tau} \Theta_A+\frac{1}{2} D_B\chi D_B\Theta_A \,.\label{eq:dbarth2}
\end{align}

In order to fully describe the boundary dynamics of regular black holes geometries we must evaluate \eqref{eq:dpsi2} and \eqref{eq:dbarth2} in $\Psi=\Psi_\ast$ determined by \eqref{eq:22}.   These configurations are invariant under a single cover of OSp$(2,2)$ for the values $n=\frac{1}{2}$ and $m=0$, therefore we choose $\Psi_\ast=\frac{\pi^2}{\beta^2}\theta_1\theta_2$.

Similarly to what was done in section \ref{bdf}, it is useful to express \eqref{onshell2} in terms of $z(T, \Theta_A)$. To achieve this, one can use the expression of the super-Schwarzian function transforming under superreparametrizations 
\begin{equation}
S^{(2)}(T, \Theta_A;z)=-\Omega_{\Theta}\, S^{(2)}(z; T, \Theta_A )
\end{equation}
such that
\begin{equation}\label{onshell3}
I_{\rm red}=\frac{k y_0}{2\pi}\int \extd T \extd\Theta_{1}\extd\Theta_{2}  \left[ \frac{\pi^2}{\beta^2} \theta_1\theta_2 \Omega_{z} +S^{(2)}(z;T, \Theta_A)\right], 
\end{equation}
where $\Omega_{z}$ is the determinant associated to the inverse of $D_A\Theta_{B}$. Thus, we have obtained that the boundary dynamics -- for bulk configurations with a nonvanishing Casimir -- turns out to be controlled by the $\cN=2$ super-Schwarzian action, whose form was first proposed in \cite{Fu:2016vas}. It has been argued in \cite{Forste:2017apw}, that through a different regularization method,  the second order formulation of $\cN=(2,2)$ dilaton supergravity yields the zero temperature version of \eqref{onshell3}.

\subsection[\texorpdfstring{$\cN=2$}{N=2} superconformal mechanics]{\texorpdfstring{$\boldsymbol{\cN=2}$}{N=2} superconformal mechanics}

Here, we will study the boundary dynamics associated to the class of configurations that possesses a vanishing Casimir function. In this case the bulk action is shown to reduce to the one of $\cN =2$ superconformal mechanics, whose space of solutions corresponds to the orbit of bulk BPS states. The latter fact is in fully agreement with the observation made in section \ref{sec:N1superconf} for the $\cN=1$ case. 

Let us proceed by writing the determinant of the extended superdiffeomorphism $\Theta_A$ as $\Omega_{\Theta}=K^ {-2}$, where $K=q-\theta_A \zeta_A+\theta_1\theta_2u$ is an even superfield of conformal dimension $-\frac{1}{2}$.
In terms of $K$, the phase space \eqref{coadN} becomes
\eq{ 
Y=K^{2} y_0 \qquad\qquad \Psi=K^{-2}( \Psi_0-K D_1D_2 K)\,.
}{eq:Rep2}
For these configurations, the variation of $I_B$ is given by 
\eq{
\delta I_{B}=\frac{k y_0}{2\pi}\intop \extd z \left[\delta \Psi_{0}-2\delta \left[\log(K)\right] \Psi_{0}\right]\,.
}{IBK2}
In order to integrate the second term it is necessary to impose an integrability condition on $\Psi_0$.  Since $\Psi_0$ stands for an even superfield, one could consider the set of configurations satisfying 
\be
\label{CBCQM}
\Psi_0\approx g\,,
\ee
with $g$ a constant without variation. From \eqref{integraN}, it can be seen that, in terms of supercoordinates $\Theta_{A}$, this means that the asymptotic value of the Casimir function \eqref{CasimirN2} vanishes and $\cP=-g$.
Within this set of solutions, the second term in \eqref{IBK2} can be easily integrated yielding 
\be
-\frac{k y_0}{\pi} g \intop \extd z  \log(K) \,.
\label{IBNK3}
\ee
Thus, the action principle $I_{\rm bulk}+I_{B}$ is well defined for configurations that satisfy \eqref{CBCQM}. We then must evaluate the action on the constraint surface, which implies that its value is only given by $I_{B}$. Using \eqref{eq:Rep2} to express $\Psi_0$ in terms of $\Psi$ and $K$, and integrating by parts, we find that
\begin{equation}\label{N2sc}
I_{\text{red}}[K;\Psi]=\frac{ky_{0}}{2\pi}\int \extd\tau \extd\theta_{1}\extd\theta_{2}\left[K^{2}\Psi-D_{1}KD_{2}K-2g \log(K)  \right]\,.
\end{equation}

The action of the reduced theory \eqref{N2sc} corresponds to the one of $\cN=2$ superconformal mechanics, that besides of being endowed with an inverse square potential, it contains a harmonic-like term with coupling constant $g$. This latter term ensures the existence of a ground state at quantum level \cite{deAlfaro:1976je,Akulov:1984uh,Fubini:1984hf}.\footnote{Note that the action \eqref{N2sc} can also be recovered in the context of nonlinear realizations of the $\cN=2$ super-Virasoro group \cite{Akulov:2004bd}. }

Due to the presence of the external source $\Psi$, the action \eqref{N2sc} is seemingly invariant under the entire SDiff$(S^1)$ group with $\cN=2$. Nonetheless, consistency with integrability condition \eqref{CBCQM} imposes that $\cL_\ast=0$, $\psi^A_\ast=0$, $\cP_\ast=-g$.  By virtue of regularity (see section \ref{HoloN2}) the latter representative is only consistent with periodic boundary conditions (taking $\Gamma_+$  in \eqref{eq:HolonomyN}, which in turn, is determined by \eqref{eq:22} with $n=m$. This fact breaks the $\cN =2$ SDiff$(S^1)$ symmetry down to a graded U$(1)$ symmetry and additionally imposes the following quantization condition on the coupling constant
\begin{equation}
g=-\frac{2\pi m}{\beta}\,.
\end{equation} 

On-shell, the boundary dynamics of \eqref{N2sc} encodes the set of configurations considered in this section. Indeed, the equation of motion that follows from the action principle reads
\begin{equation}
K\Psi_\ast+D_{1}D_{2}K+\frac{2\pi m}{\beta} K^{-1}=0
\end{equation}
which is nothing but condition \eqref{CBCQM}.  Thus, as anticipated at the beginning of this subsection, solutions of the  boundary model \eqref{N2sc} are mapped into ground states of the bulk theory, that according to \cite{Bergamin:2003mh} correspond to bulk BPS-like states.

\section{Particle on a supergroup manifold}\label{ultimasec}

So far, we have successfully determined the boundary dynamics associated to $\cN=1$ and $\cN=2$ dilaton supergravity equipped with the set of asymptotic conditions introduced in sections \ref{sec:fo} and \ref{sec:Asymptotic conditions 2}, respectively. The crucial point in these derivations  has been to find consistent integrability conditions between the fields $\textbf{X}$ and $\textbf{A}$ such that the boundary term required for the variational principle,
\begin{equation}
\delta I_{B}=-\frac{k}{2\pi}\ointop \extd\tau\,\text{Str}\left[\textbf{X\ensuremath{\delta}}\textbf{A}_{\tau}\right]\label{eq:dIBultimo}
\end{equation}
yields a local expression for $I_B$. In order to deal with the integrability of \eqref{eq:dIBultimo} for $\cN >2$, we will consider the set of asymptotic conditions consistent with the loop group of OSp$(2,N)$, and then impose an appropriate boundary condition between $\mathbf{A}$ and $\mathbf{X}$. It will be also shown that the above treatment allows to reduce the dynamics of the theory to the one of a particle moving in a group manifold.  

In a second stage, we restrict ourselves to asymptotically AdS$_2$ supergeometries \eqref{AXExt2} containing a fluctuating dilaton multiplet. This is achieved by applying a suitable Hamiltonian reduction that yields some constraints on the particle model. Finally, we consider geometries representing the orbit of regular black hole solutions. These reductions not  only agree with the results obtained in the previous sections for $N=1$ and $N=2$ cases but also extend them to $N>2$, which lead to a generic class of super-Schwarzian actions for generic OSp$(2,N)$ dilaton supergravity models. 

\subsection{Integrability condition and regularized action principle}\label{sec:integ}

We parameterize our asymptotic conditions as in \eqref{AXExt1}, 
\be
\label{loop}
\textbf{A}=b^{-1}(d+a)b+\cO(r^{-2}) \qquad\qquad \textbf{X}=b^{-1}x b +\cO(r^{-2}).
\ee
We assume that the radial-independent fields $a$ and $x$ satisfy the following integrability condition
\be
\label{prop}
{a}_{\tau}=f_{\tau} {x}
\ee
with $f_{\tau}$ an arbitrary function that is assumed to play the role of a one-form component. It must be noted that the integrability condition \eqref{prop} has been previously used for JT gravity \cite{Brigante:2002rv} and for suitable bosonic extensions thereof \cite{Mertens:2018fds, Gonzalez:2018enk}.

The solution space of ${x}$ constraining by \eqref{prop} reduces the dilaton equation $\delta_{\textbf{X}}\textbf{A}=0$ to $\d_{\tau} {x}=0$. Combining the above to the fact that the gauge connection is flat ${a}_{\tau}=g^{-1}\d_{\tau}g$, we find that the group element $g$ obeys the following condition
\be
\label{yapo}
\d_{\tau}(f_{\tau}^{-1} g^{-1} \d_{\tau} g)=0
\ee
which turns out to be the equation of motion for a particle in a group manifold. The variational principle that leads to equations \eqref{yapo} will be obtained by regularizing the BF model through the integration of \eqref{eq:dIBultimo}. For this, we make use of the asymptotic conditions \eqref{loop} together with \eqref{prop}. Thus, by assuming that $f_\tau$ has a fixed zero mode, it follows that the integrated expression for $I_B$ is given by
\eq{ 
I_{B}=-\frac{k}{4\pi}\ointop \extd\tau \; f_{\tau} \, {\rm Str}({x}^2)\,.
}{eq:IBN}

The integrability conditions (\ref{prop}) play the same role as boundary conditions on canonical variables. In a theory with $n$ canonical pairs $(q^i,p_i)$ one has to impose $n$ boundary conditions to achieve a consistent variation principle. Here, we also imposed $n$ conditions on $n$ pairs of boundary variables $(a,x)$, but we managed to do better: the conditions (\ref{prop}) contain an arbitrary function $f_\tau$ that fluctuates freely except for the zero mode.

We recall that the full action principle consists of the BF bulk piece \eqref{eq:BFaction} plus the boundary term $I_B$ \eqref{eq:IBN}. Nonetheless, as shown in section \ref{bdf}, once the theory is evaluated on the constraint surface, the action becomes the boundary term $I_B$ by virtue of the asymptotic conditions \eqref{loop}. Therefore, from \eqref{prop} one obtains that the action reduces to
\be
I_{\rm red}[g]=-\frac{ky_0}{4\pi} \ointop \extd f \,\text{Str}\left[(g^{-1}\partial_{f}g)^2\right]\,.\label{eq:Igeom}
\ee
Here, we have used that $f_{\tau}=\frac{1}{y_0} \d_{\tau} f$ ensuring that $f_{\tau}$ has a fixed zero mode. The set of symmetries, making  $I_{\rm red}[g]$ invariant, corresponds to left and right 
 multiplications
\be
\label{sym}
g \mapsto  h_L \, g \, h_R^{-1}
\ee
where $h_L$ and $h_R$ are constant elements of OSp$(2,N)$. The action on the right is the residual symmetry subjects to the integrability condition \eqref{prop}. On the other hand, left symmetry appears after imposing that ${a}_{\tau}\extd\tau$ is a flat connection defined by a left invariant Maurer-Cartan form. In the next section we will see how 
the right symmetry
is lost after restricting the phase space to asymptotically AdS spaces, and that the left symmetry, which is the only one that persists, implements the OSp$(2,N)$ symmetry in the super-Schwarzian theory at the boundary.

\subsection{Hamiltonian reduction: extended super-Schwarzian action}

The purpose of this subsection is to constrict the configurations to be asymptotically AdS$_2$ supergeometries, which results in the following form of the gauge field
\begin{equation}
{a}=\left[L_{1}+\left( {\cL}+\frac{1}{2} {\cP}^{AB} {\cP}^{AB}\right)L_{-1}+{\psi}^{A}G_{-\frac{1}{2}}^{A}+\frac{1}{2} {\cP}^{AB}T^{AB}\right] \extd f \,.\label{eq:Ext-at}
\end{equation}
Subsequently, the Hamiltonian reduction is performed by considering the Gauss decomposition of an arbitrary OSp$(2,N)$
group element\footnote{%
See appendix \ref{sec:A1} for an explicit matrix representation of $g$.}
\begin{equation}
\label{gauss}
g=\exp\big[{h(f)L_{1}+\eta_{A}(f)G_{\frac{1}{2}}^{A}}\big] \exp\big[2q(f)L_{0} + \frac{1}{2}\sigma_{AB}(f)T^{AB}\big] \exp\big[{k(f)L_{-1}+\zeta_{A}(f)G_{-\frac{1}{2}}^{A}}\big]
\end{equation}
whose associated Maurer-Cartan form is assumed to be equal to the gauge field encoding the AdS$_2$ supergeometries \eqref{eq:Ext-at}, i.e.~${a}_{f}=g^{-1}\d_{f}g$. This equality leads to the set of relations
\begin{align}
{\cL}& = -q\text{\ensuremath{''}}-(q\text{\ensuremath{'}})^{2}+\zeta_{B}\zeta_{B}\text{\ensuremath{'}}-\frac{1}{2}(M_{AB}\text{\ensuremath{'}})^2\label{L}\\
{\cP}_{AB} & = M_{CA}M_{CB}\text{\ensuremath{'}}+\zeta_{A}\zeta_{B}\\
{\psi}_{A} & = \zeta_{A}\text{\ensuremath{'}}+q\text{\ensuremath{'}}\zeta_{A}+M_{CA}M_{CB}\text{\ensuremath{'}}\zeta_{B}
\end{align}
where
\begin{equation}\label{relations}
\zeta_{A}=-e^{q}\eta_{B}\text{\ensuremath{'}}M_{BA}\qquad k=-q'\qquad e^{-2 q}=h\text{\ensuremath{'}}+\eta_{B}\eta_{B}\text{\ensuremath{'}}\qquad M=e^{\frac{1}{2}\sigma_{AB}m_{AB}}
\end{equation}
with $ \left(m^{AB}\right)_{\,\,\,D}^{C}=2\delta^{C[A}\delta_{D}^{B]}$ and primes denote differentiation with respect to $f$.  At this point, one can notice that the condition 
$g^{-1}\d_{f}g= L_1+\cdots$ fixes the right symmetry ($h_R$)  to be generated only by $L_{1}$. One can completely fix this residual freedom by absorbing $h_R$ in a redefinition of the function $h(f)$ appearing in \eqref{gauss} and thus the right action is not a symmetry of the reduced action principle anymore. 

By replacing \eqref{eq:Ext-at} in \eqref{eq:Igeom}, the reduced action becomes
\begin{equation}
I_{\text{red}} = \frac{ky_0}{2\pi}\int \extd f \,\cL\,.
\end{equation}
Then, we proceed to compute $\cL$ in terms of the fields
defining the Gauss decomposition \eqref{gauss}. One can readily see from \eqref{L} that  in terms of the field $(q,\zeta_{A},M_{AB})$ the reduced model corresponds to the action of a free superparticle related to the coadjoint orbit passing through $\cL_\ast=\cP^{AB}_\ast=\psi^{A}_\ast=0$. On the other hand, solving the constraints \eqref{relations} in terms of the variables $(h,\eta_{A},M_{AB})$ we are able to find a generalization of the super-Schwarzian action. In this case,   \eqref{L} takes the following form 
\begin{equation}
\label{tilL}
2\mathcal{L}= \frac{\left(h''+\eta_{B}\eta_{B}''\right)'}{h'+\eta_{A}\eta_{A}'}-\frac{3}{2}\left(\frac{h''+\eta_{B}\eta_{B}''}{h'+\eta_{A}\eta_{A}'}\right)^{2}+\frac{2\eta_{B}'\eta_{B}''}{h'+\eta_{A}\eta_{A}'}+\frac{2\eta_{B}'\eta_{C}'}{h'+\eta_{A}\eta_{A}''}M_{BE}M_{CE}'-(M_{AB}')^2
\end{equation}
which turns out to be an explicit representation of the extended super-Schwarzian as it is invariant under  global OSp$(2,N)$ transformations inherited from the left action \eqref{sym}. Infinitesimally, this corresponds to $\delta g = \lambda_{L} g$ that in terms of the fields $ h,\eta_{A},M_{AB}  $ reads
\begin{align}
\delta h & =  \xi_{+}+\xi_{-}h+\xi_{0}h^{2}-\epsilon_{-}^{A}\eta_{A}h+\epsilon_{+}^{A}\eta_{A}\\
\delta\eta_{A} & =  \frac{1}{2}\xi_{-}\eta_{A}+\xi_{0}\eta_{A}h+\epsilon_{-}^{A}h-\epsilon_{-}^{B}\eta_{B}\eta_{A}-\epsilon_{+}^{A}-\frac{1}{2}\rho_{AB}\eta_{B}\\
\delta M_{AB} & =  -\xi_{0}\eta_{A}\eta_{C}M_{CB}+\epsilon_{-}^{C}\eta_{A}M_{CB}-\epsilon_{-}^{A}\eta_{C}M_{CB}-\frac{1}{2}\rho_{AC}M_{CB}
\end{align}
with Lie-algebra-parameter $\lambda_{L} =\xi_{m}L_{m}+\epsilon_{p}^{A}G_{p}^{A}+\frac{1}{2}\rho_{AB}T_{AB}$. 

Until now we have constructed the super-Schwarzian theory at zero temperature. To introduce finite temperature, we need to provide a map in such a way that the connection $a$ in \eqref{eq:Ext-at}  lie on the orbit of configurations with trivial holonomy,  $\cL_\ast=\frac{\pi^2}{\beta^2}$ and 
$\psi^A_\ast=\cP^{AB}_\ast=0$. 
This restriction is ensured by introducing new fields $\theta(f)$ and $\nu_{A}(f)$ through the following map
\be
\label{temp}
h(f)= \tan \big(  \theta (f)/2\big) \qquad\qquad \eta_{A}(f)=\sqrt{h'(f)}\, \nu_{A}(f)
\ee
where $\nu_{A}(f+2\pi)=\nu_A(f)$ and $\theta$ satisfies $\theta(f+2\pi)=\theta(f)+2\pi$ (for definiteness we assume that $\beta=2\pi$). We can prove that this is the right map by studying the perturbations of the Lagrangian \eqref{tilL} around the saddle solution 
\be
\label{pert}
\theta(f)= f+ \varepsilon(f)\qquad\qquad \nu_A(f)\qquad\qquad M_{AB}= \delta_{AB}+\sigma_{AB}(f)
\ee
with $\varepsilon$, $\nu_A$ and $\sigma_{AB}$ small deviations.  Keeping up to quadratic terms, we find
\begin{multline}
\cL= \frac{1}{4}+\frac{1}{2} (\ve' +\ve''')+ \frac{1}{4 }(\ve')^2-\frac{1}{2}\ve'\ve'''-\frac{3}{4}(\ve'')^2 \\ + \frac{3}{2}\nu_A' \nu_{A}'' + \frac{1}{2}\nu_A \nu_{A}''' - \frac{1}{4} \nu_{A}\nu_{A}' - \frac{1}{2}(\sigma_{AB}')^2 + \cdots\,.
\end{multline}
Recalling that $\beta=2\pi$, the above relation shows that \eqref{pert} is a perturbation of the point $\cL= \pi^2/\beta^2$ . Furthermore, they vanish for $\mathfrak{osp}(2,N)$ transformations 
\be
\ve=\ve_{n} e^{in f} \qquad\qquad \nu_{A}=\nu_A^{p} e^{i p f} \qquad\qquad \sigma_{AB}=\sigma^0_{AB}
\ee
with $n=0, \pm1$, $p=\pm\frac{1}{2}$ and $\sigma^0_{AB}$ a constant $\mathfrak{so}(N)$ Lie algebra element. This last step tells us that map \eqref{temp} brings the system to the orbit associated to the Neveu-Schwarz sector [$\Gamma_{-}$ for $N=1$ and $N=2$ in \eqref{eq:HolonomyN}]. 

\section{Beyond the highest-weight ansatz}\label{se:9}

Usually, the gauge connection $a$ is considered to have highest-weight form
\begin{equation}
a=L_{1}+Q\label{m01}
\end{equation} 
where
\begin{equation}
\left[L_{-1},Q\right]=0\,.\label{m02}
\end{equation}
For example, it can be easily shown that the connections \eqref{eq:A} and \eqref{AXExt2} fulfill the highest-weight condition \eqref{m02}. The main motivation for this choice is that the leading terms in the metric are fixed, which simplifies the analysis considerably.

In \cite{Grumiller:2017qao} more general  boundary conditions for JT gravity were considered that went beyond the highest weight ansatz  allowing all fields in the $\mathfrak{sl}\left(2\right)$ basis to fluctuate.  In this section, we show that one can consider more general sets of $\mathrm{AdS}_2$ boundary conditions also in the supergravity context.

Note, that the construction of the boundary action and the integrability conditions of section \ref{sec:integ} do not rely on the highest-weight ansatz and thus can be used to address the consistency of the variation principle for the boundary conditions described below.

In this section, we shall only analyze the asymptotic symmetries. The full analysis of boundary theories shall be postponed to some future work.

\subsection{Some field redefinitions}
Our strategy is quite simple. First we define for each set of boundary conditions the set of large gauge transformations that preserve the form of asymptotic fields. Then we make some redefinitions of the fields and the transformation parameters, so that the action of asymptotic symmetries coincides with a representation of a known algebra. This algebra is then identified with the asymptotic symmetry transformations for that boundary conditions. Of course, the success of this procedure depends on our ability to find the right redefinitions. The method is thus non-generic. It works for the cases that we list below. In this subsection we present some general redefinitions that we write for any $\mathfrak{osp}(2,N)$ model. 

For our purposes it is enough to analyze the connection variables only. Let us write the asymptotic field $a$ [see \eqref{loop}] in components  as 
\begin{equation}
a =	\big[\mathcal{L}^{m}(\tau)L_{m}+\psi_{A}^{\alpha}(\tau)G_{\alpha}^{A}+\frac{1}{2}\mathcal{S}_{AB}(\tau)T_{AB}\big]\,\mathrm{d}\tau\,.
\label{m03}
\end{equation}
Note that in contrast to \eqref{eq:A} and \eqref{AXExt2}, all fields in the $\mathfrak{osp}(2,N)$ basis of the gauge connection \eqref{m03} are allowed to vary.

Now, let us rescale the fermionic fields, $\psi_{A}^{+}\equiv\left(\mathcal{L}^{+}\right)^{-1/2}\psi_{A}^{\frac{1}{2}}$, $\psi_{A}^{-}\equiv\left(\mathcal{L}^{+}
\right)^{1/2}\psi_{A}^{-\frac{1}{2}}$ and define
\begin{align}
\psi_{A} &\equiv	\psi_{A}^{-}-\psi_{A}^{+}{}'-\mathcal{J}\psi_{A}^{+}
-\mathcal{P}_{AC}\psi_{C}^{+}\label{m08a}\\
\mathcal{P}_{AB} &\equiv \mathcal{S}_{AB}+\psi_{A}^{+}
\psi_{B}^{+}\label{m08b}\\
\mathcal{L}	&\equiv	\mathcal{T}-\mathcal{J}^{2}-\mathcal{J}'-2\psi_{A}^{+}\psi_{A}^{-}+\psi_{A}^{+}\psi_{A}^{+}{}'+\mathcal{P}_{AB}\psi_{A}^{+}\psi_{B}^{+}-\frac{1}{2}\mathcal{P}_{AB}\mathcal{P}_{AB}\label{m08c}
\end{align}
where
\begin{equation}
\mathcal{T}\equiv\mathcal{L}^{+}\mathcal{L}^{-}\qquad\qquad\mathcal{J}\equiv-\frac12\,\partial_\tau\ln{\mathcal{L}^+} + \frac{1}{2}\mathcal{L}^{0}\,. \label{m09}   
\end{equation}
In the absence of fermions and SO$(N)$ gauge fields, $\mathcal{L}$ in \eqref{m08c} reduces to the mass function defined in \cite{Grumiller:2017qao}.

We also write in components the parameter of asymptotic symmetry transformations
\begin{equation}
\lambda	=	\xi^{m}L_{m}+\epsilon_{A}^{\alpha}G_{\alpha}^{A}+\frac{1}{2}\nu_{AB}T_{AB}\label{m11}
\end{equation}
that is assumed to depend on $\tau$ only. The components of $a$ are transformed as
\begin{align}
\delta_\lambda \mathcal{L}^{0}	&=	\mathcal{\xi}^{0}{}'+2\mathcal{\mathcal{L}}^{+}\xi^{-}-2\mathcal{\mathcal{L}}^{-}\xi^{+}-2\psi_{A}^{\frac{1}{2}}\epsilon_{A}^{-\frac{1}{2}}-2\psi_{A}^{-\frac{1}{2}}\epsilon_{A}^{\frac{1}{2}}\label{m12a}\\
\delta_\lambda \mathcal{L}^{\pm}	&=	\xi^{\pm}{}'\mp\mathcal{\mathcal{L}}^{0}\xi^{\pm}\pm\mathcal{\mathcal{L}}^{\pm}\xi^{0}-2\psi_{A}^{\pm\frac{1}{2}}\epsilon_{A}^{\pm\frac{1}{2}}\label{m12b}\\
\delta_\lambda \psi_{A}^{\pm\frac{1}{2}}	&=	\epsilon_{A}^{\pm\frac{1}{2}}{}'\mp\frac{1}{2}\mathcal{\mathcal{L}}^{0}\epsilon_{A}^{\pm\frac{1}{2}}\pm\mathcal{\mathcal{L}}^{\pm}\epsilon_{A}^{\mp\frac{1}{2}}\pm\frac{1}{2}\xi^{0}\psi_{A}^{\pm\frac{1}{2}}\mp\xi^{\pm}\psi_{A}^{\mp\frac{1}{2}}+\mathcal{P}_{AB}\epsilon_{B}^{\pm\frac{1}{2}}-\nu_{AB}\psi_{B}^{\pm\frac{1}{2}}\label{m12c}\\
\delta_\lambda \mathcal{P}_{AB}	&=	\nu'_{AB}-\psi_{A}^{\frac{1}{2}}\epsilon_{B}^{-\frac{1}{2}}+\psi_{B}^{\frac{1}{2}}\epsilon_{A}^{-\frac{1}{2}}+\psi_{A}^{-\frac{1}{2}}\epsilon_{B}^{\frac{1}{2}}-\psi_{B}^{-\frac{1}{2}}\epsilon_{A}^{\frac{1}{2}}+\mathcal{P}_{AC}\nu_{CB}-\mathcal{P}_{BC}\nu_{CA}\,.\label{m12d}
\end{align}
It is useful to introduce new transformation parameters
\begin{align}
\xi	&= \left(\mathcal{L}^{+}\right)^{-1}\xi^{+}\\
\epsilon_A &= \left(\mathcal{L}^{+}\right)^{-1/2}\epsilon^{\frac{1}{2}}_{A}-\psi_{A}^{+}\xi\\
\mu_{AB} &= \nu_{AB}-\xi\mathcal{S}_{AB}+\psi_{A}^{+}\epsilon_{B}-\psi_{B}^{+}\epsilon_{A}\,.\label{e08}
\end{align}
After these redefinitions we reproduce the transformation rules \eqref{dL}-\eqref{dP} for quantities $\mathcal{L}$, $\psi_A$, $\mathcal P_{AB}$ for the highest-weight asymptotic conditions. 

\subsection[Asymptotic symmetries for \texorpdfstring{$\mathfrak{osp}(2,1)\;\mathrm{BF}$}{osp(2,1) BF} gravity]{Asymptotic symmetries for \texorpdfstring{$\boldsymbol{\mathfrak{osp}(2,1)\;\mathrm{BF}}$}{osp(2,1) BF} gravity}\label{se:9.1}

\subsubsection{Loop group boundary condition}\label{sec:loop1}
Our first example corresponds to the case when all components of the boundary connection $a$ are allowed to vary. This form of boundary conditions is preserved by arbitrary gauge transformations with the parameters $\xi^m(\tau)$, $\epsilon^\alpha(\tau)$ that depend just on the boundary coordinate $\tau$. The corresponding asymptotic algebra is just the algebra of one-dimensional $\mathfrak{osp}(2,1)$ gauge transformations or the current algebra of $\mathfrak{osp}(2,1)$. This very simple example has a rather important property. The asymptotic symmetry algebra obtained contains two independent one-dimensional supergenerators, i.e. we have an $\mathcal{N}=2$ boundary supersymmetry. Unfortunately, we were not able to get such an extended supersymmetry in less trivial examples.

\subsubsection{Superconformal boundary condition}
Alternatively, let us consider the boundary conditions $\mathcal{L}^{0}=0$, $\mathcal{L}^{+}=1$ and $\psi^{\frac{1}{2}}=0$ in  \eqref{m08a} and \eqref{m08c}. In terms of $\mathcal L = \mathcal L^-$ and $\psi = \psi^-$ the gauge connection \eqref{m03} reduces to \eqref{eq:A}. 

The gauge parameter preserving these boundary conditions is
\begin{equation}
\lambda = \xi L_1 -\xi'L_0 +  \big(\frac{1}{2}\xi''+\mathcal{L}\xi+\psi\epsilon\big)\, L_{-1} + \epsilon G_{\frac{1}{2}} + \big(-\epsilon'+\xi\psi\big)\,G_{\frac{1}{2}} \,.
\end{equation}
The transformations \eqref{m12b}, \eqref{m12c} of $\mathcal L$ and $\psi$ take the form \eqref{eq:deltaL} and \eqref{eq:deltapsi}, thus giving the superconformal symmetry algebra that has been considered in detail above.

\subsubsection{Warped superconformal boundary condition}
Let us set $\mathcal{L}^{0}=\psi^{\frac{1}{2}}=0$ while allowing $\mathcal{L}^{\pm}$ and $\psi^-\equiv \psi$ to vary. Through the conditions $\delta\mathcal{L}^{0}=\delta\psi^{\frac{1}{2}}=0$, we can express the parameters $\xi^-$ and $\epsilon^{-\frac{1}{2}}$ in terms of $\xi^0$, $\xi$ and $\epsilon$ 
\begin{align}
\xi^{-}	&=	\big(\mathcal{L}^{+}\big)^{-1}\big[-\frac{1}{2}\xi^{0}{}'+\big(\mathcal L + \mathcal{J}'+\mathcal{J}^2\big)\xi+\mathcal{\psi}\epsilon\big]\\
\epsilon^{-\frac{1}{2}}	&= \big(\mathcal{L}^{+}\big)^{-1\text{/2}}\big(-\epsilon'+\mathcal{J}\epsilon+\mathcal{\psi}\xi\big)\,.
\end{align}
We have the following transformation rules
\begin{align}
\delta\mathcal{J}	&=	\mathcal{J}\xi'+\mathcal{J}'\xi-\frac{1}{2}\xi^{0}{}'-\frac{1}{2}\xi''\label{trJ}\\
\delta\mathcal{T}	&=	2\mathcal{T}\xi\text{\ensuremath{'}}+\mathcal{T}\text{\ensuremath{'}}\xi-\mathcal{J}\,\xi^0{}'-\frac12\,\xi^0{}''+3\psi\epsilon\text{\ensuremath{'}}+\psi\text{\ensuremath{'}}\epsilon\\
\delta \psi & =  -\epsilon''-\mathcal{L}\epsilon+\frac{3}{2}\psi\xi'+\psi'\xi\, .
\end{align}
Note that $\mathcal J$ is unaffected by the supersymmetry, and $\psi$ transforms as \eqref{eq:deltapsi}, while $\mathcal{L}=\mathcal{T}-\mathcal{J}^2-\mathcal{J}'$ transforms exactly as in \eqref{eq:deltaL}. The algebra of $\mathcal J, \mathcal T, \psi$ corresponds to the warped superconformal algebra (see e.g.~\cite{Nakayama:2013coa}) with a twist term. For vanishing fermions it reduces to the algebra (6.28) in \cite{Grumiller:2017qao}. 

Using as generators Fourier modes for the functions $\mathcal{J, L, \psi}$ with normalizations $L_n=k/(2\pi)\oint e^{in\tau}\mathcal{L}\,\extd\tau$ (and similarly for $J_n$ and $\psi_n$), shifting $\lambda:=\xi^0 + \xi^\prime$, and converting the transformation behavior above into commutation relations yields the (untwisted) warped superconformal algebra,
\begin{align}
    [L_n,\,L_m] &= (n-m)\,L_{n+m} + \frac{k}{2}\,n^3\,\delta_{n+m,\,0} \\
    [L_n,\,\psi_m] &= (n/2-m)\,\psi_{n+m} \\
    [L_n,\,J_m] &= -m\,J_{n+m} \\
    \{\psi_n,\,\psi_m\} &= -L_{n+m} + i k\, n^2\,\delta_{n+m,\,0} \\
    [\psi_n,\,J_m] &= 0 \\
    [J_n,\,J_m] &= \frac k2\, n\,\delta_{n+m,\,0}\,.
\end{align}

\subsubsection[All \texorpdfstring{$\mathcal{L}^m$}{Lm} vary while \texorpdfstring{$\psi^{\frac{1}{2}}=0$}{psi1/2=0}]{All \texorpdfstring{$\boldsymbol{\mathcal{L}^m}$}{Lm} vary while \texorpdfstring{$\boldsymbol{\psi^{\frac{1}{2}}=0}$}{psi1/2=0}}

The condition $\delta \psi^{\frac{1}{2}}=0$ leads to an expression for $\xi^+$ through $\xi$ and $\sigma$,
\begin{equation}
\epsilon^{-\frac{1}{2}}=\left(\mathcal{L}^{+}\right)^{-1\text{/2}}\left(-\epsilon'+\mathcal{J}\epsilon+\mathcal{\psi}\xi\right).
 \end{equation}
Thus, we have the following transformation rules for gauges fields
\begin{align}
\delta\mathcal{L}^{0} &= \xi^{0}{}'+2\alpha-2\mathcal{T}\xi-2\mathcal{\psi}\epsilon \\
\delta\mathcal{J} &= -\frac{1}{2}\xi''+\mathcal{J}\xi'+\alpha+\left(-\mathcal{T}+\mathcal{J}'\right)\xi-\mathcal{\psi}\epsilon \\
\delta\mathcal{T} &= \mathcal{T}\xi'-2\mathcal{T}\mathcal{J}\xi+\alpha'+2\mathcal{J}\alpha+2\mathcal{\psi}\epsilon'
-2\mathcal{J}\mathcal{\psi}\epsilon \\
\delta\psi	&=	-\epsilon''-\mathcal{L}\epsilon+\frac{3}{2}\xi'\psi+\mathcal{\psi}'\xi 
\end{align}
where $\alpha= \mathcal L^+ \xi^{-}$.

To remove non-linearities in $\delta \mathcal T$  we redefine the parameter $\alpha$ as $\alpha=-\frac{1}{2}\lambda'+\mathcal{T}\xi+\psi\epsilon$, yielding
\begin{align}
\delta\mathcal{J} &= \mathcal{J}\xi'+\mathcal{J}'\xi-\frac{1}{2}\lambda'-\frac{1}{2}\xi''\\
\delta\mathcal{T} &=	2\mathcal{T}\xi\text{\ensuremath{'}}+\mathcal{T}\text{\ensuremath{'}}\xi-\mathcal{J}\lambda'-\frac12\lambda''+3\psi\epsilon\text{\ensuremath{'}}+\psi\text{\ensuremath{'}}\epsilon\\
\delta\psi &=-\epsilon''-\mathcal{L}\epsilon+\frac{3}{2}\xi'\psi+\psi'\xi.
\end{align}
The parameter $\xi^0$ only appears in $\delta \mathcal L^0$ which now reads
\begin{equation}
\delta\mathcal{L}^{0} =\xi^{0}{}'-\lambda'.
\end{equation}
The algebra is linear and closed on $(\mathcal{J},\mathcal{T},\mathcal{L}^0,\psi)$. It is an $\mathcal{N}=1$ super Virasoro plus a $\mathfrak{u}(1)$ current algebra plus translations on $\mathcal{L}^0$. Except for the translations, this algebra is the same as we had in the previous section for the warped conformal case.

\subsection{Asymptotic symmetries for \texorpdfstring{$\mathfrak{osp}(2,2)$}{osp(2,2)} BF gravity}\label{se:9.2}

\subsubsection{Loop group boundary condition}
By allowing all components in the connection to fluctuate, exactly as we did in section \ref{sec:loop1} for $N=1$, one obtains that the current algebra of $\mathfrak{osp}(2,2)$ plays the role of an asymptotic symmetry algebra. Again, the doubled number of one-dimensional supersymmetries is the main lesson here.

\subsubsection{Extended superconformal boundary condition} 
Let us impose the boundary conditions $\mathcal L^0=0$, $\mathcal L^+=1$ and $\psi^{\frac{1}{2}}_A=0$. Stability of these conditions under the gauge transformations implies the following relations on the gauge parameter components
\begin{align}
\xi^{0} &= -\xi'\\
\xi^- &= \frac{1}{2}\xi''+\big(\mathcal L + \frac{1}{2}\mathcal P_{AB}\mathcal P_{AB}\big)\xi+\psi_{A}\epsilon_{A}\\
\epsilon_{A}^{-\frac{1}{2}} &= \psi_{A}\xi-\epsilon'_{A}-\mathcal{P}_{AB}\epsilon_{B}
\end{align}
so that
\begin{eqnarray}
\delta \mathcal{L} &=& \frac{1}{2}\xi'''+2\xi'\mathcal{L}+\mathcal{L}'\xi+3\psi_{A}\epsilon'_{A}+\psi'_{A}\epsilon_{A}-\mathcal{P}_{AB}\mu'_{AB}\\
\delta \psi_A &=&	\frac{3}{2}\xi'\psi_{A}+\xi\psi'_{A}-\mu_{AB}\psi_{B}-\epsilon''_{A}-\mathcal{L}\epsilon_{A}-\mathcal{P}'_{AB}\epsilon_{B}-2\mathcal{P}_{AB}\epsilon'_{B} \label{dp02}\\
\delta \mathcal P_{AB} &=& \mu'_{AB}+\xi'\mathcal{P}_{AB}+\mathcal{P}'_{AB}\xi+2\mathcal{P}_{C[A}\mu_{B]C}+2\psi_{[A}\epsilon_{B]}.
\end{eqnarray}
Equations \eqref{dp02} and \eqref{dp} coincide for $N=2$. Furthermore, $\mathcal L, \psi_A$ and $\mathcal P_{AB}$ generate one copy of the SO$(2)$-extended super-Virasoro algebra.

\subsubsection{Extended warped superconformal boundary condition} 
Now, we relax the restriction on $\mathcal{L}^+$ and require $\mathcal L^0=\psi^{\frac{1}{2}}=0$ leaving $\mathcal{L}$ as defined in \eqref{m08c} with \eqref{m09}, $\mathcal{J}=-\tfrac12\,\partial_\tau\ln\mathcal{L}^+$, $\psi_A = \psi^-_A$ and $\mathcal P_{AB} = \mathcal S_{AB}$ to vary on the boundary.
The above conditions imply  the following relations on the parameters
\begin{align} 
\xi^- &= (\mathcal L^+)^{-1}\big[-\frac{1}{2}\xi^{0}{'}+\big(\mathcal L + \mathcal J'+\mathcal J^2 + \frac{1}{2}\mathcal P_{AB}\mathcal P_{AB}\big)\xi+\psi_{A}\epsilon_{A}\big]\\
\epsilon_{A}^{-\frac{1}{2}} &= (\mathcal L^+)^{-1/2} \big[ \psi_{A}\xi-\epsilon'_{A}+\mathcal{J}\epsilon_{A}-\mathcal{P}_{AB}\epsilon_{B}\big] \,.
\end{align}
 The transformation of $\mathcal J$, $\mathcal L$ and $\psi_A$ read (using again $\lambda:=\xi^0+\xi'$)
\begin{align}
\delta\mathcal{J} &= -\frac{1}{2}\lambda^\prime+\mathcal{J}\xi'+\mathcal{J}'\xi\\
\delta \mathcal{L} &= \frac{1}{2}\xi'''+2\xi'\mathcal{L}+\mathcal{L}'\xi+3\psi_{A}\epsilon'_{A}+\psi'_{A}\epsilon_{A}-\mathcal{P}_{AB}\mu'_{AB}\\
\delta \psi_A &=	\frac{3}{2}\xi'\psi_{A}+\xi\psi'_{A}-\mu_{AB}\psi_{B}-\epsilon''_{A}-\mathcal{L}\epsilon_{A}-\mathcal{P}'_{AB}\epsilon_{B}-2\mathcal{P}_{AB}\epsilon'_{B} \\
\delta \mathcal P_{AB} &= \mu'_{AB}+\xi'\mathcal{P}_{AB}+\mathcal{P}'_{AB}\xi+2\mathcal{P}_{C[A}\mu_{B]C}+2\psi_{[A}\epsilon_{B]}\,.
\end{align}
The transformation of $\mathcal J$ is equal to the one for the warped superconformal case \eqref{trJ}, meaning that $\mathcal J$ is also unaffected by the SO$(2)$ gauge transformations. The algebra of $\mathcal J, \mathcal L, \mathcal \psi_A, \mathcal P_{AB}$ corresponds to the extended warped superconformal algebra.

\section{Concluding remarks}\label{CR}

We have provided a consistent set of asymptotic conditions for two-dimensional dilaton supergravity  preserved by the SO$(N)$-extended group of reparametrizations of the supercircle. This fall-off was shown to be relaxed enough in order to permit the fluctuation of the dilaton multiplet without spoiling the superreparametrization symmetry at infinity. Through a detailed analysis of the asymptotic dynamics of the model, we have established that the boundary theory strongly depends on the coadjoint orbits associated to the super-Virasoro algebra. When considering orbits associated to black hole solutions, the boundary field corresponds to a superdiffeomorphism, whose dynamics is controlled by the extended super-Schwarzian theory. On the other hand, the orbits where the representative vanishes --- as BPS states --- lead to a dynamics described by (super)conformal quantum mechanics for $\cN \leq 2$, and the extended free superparticle for $\cN>2$. The results are summarized in table \ref{tab:1}.

\begin{table}[!htbp]
\vspace{3mm}
\begin{center}
\noindent
\begin{tabular}{|c|c|c|}
\hline
{\bf  Dilaton supergravity theory} & {\bf Boundary fields} & {\bf Boundary theory}    \\ \hline
\multirow{2}{*}{$\mathfrak{osp}(2,1)$--BF} & $\Theta$ & $\cN=1$ super-Schwarzian    \\ \cline{2-3} 
                                                                    & $K$ & $\cN=1$ free superparticle     \\ \hline
\multirow{2}{*}{$\mathfrak{osp}(2,2)$--BF} &  $\Theta_A$ & $\cN=2$ super-Schwarzian    \\ \cline{2-3} 
                                                                    & $K$  & $\cN=2$ SCQM  \\ \hline
\multirow{2}{*}{$\mathfrak{osp}(2,N)$--BF} &  $(h,\eta_A,M_{AB})$ & $\cN=N$ super-Schwarzian      \\ \cline{2-3} 
                                                                    & $(q,\zeta_A,M_{AB})$  & $\cN=N$ free superparticle   \\ \hline
\end{tabular}
\end{center}
\caption{Boundary dynamics for (extended) dilaton supergravities in 2D.}
\label{tab:1}
\vspace{3mm}
\end{table}

These are the first steps towards the understanding of the relation between quantum dilaton supergravity and its corresponding supersymmetric dual. In connection with recent recent results, we would like to extend the analysis of  \cite{Blommaert:2018oro,Lin:2018xkj} by tracing the relation between Wilson lines and entanglement entropy in the supersymmetric BF formulation with the corresponding boundary observables. Furthermore, it would be interesting to understand the consequences of the different asymptotic dynamics in the factorization problem of quantum dilaton supergravity. One important step here is to expand the analysis of \cite{Harlow:2018tqv} to the supersymmetric case with Lorentzian signature. 

As shown along this work, the group structure of the asymptotic symmetries associated to the theory allows to determine in a consistent manner its boundary dynamics. Thus, for instance, in view of the results obtained in \cite{Grumiller:2015vaa, Grumiller:2016dbn}, it would be interesting to explore whether it is possible to obtain a reparametrization theory from generalized models of dilaton (super)gravity \cite{Grumiller:2002nm}. Another interesting aspect to explore is related to non-perturbative local quantum triviality of bosonic \cite{Kummer:1996hy} and supersymmetric \cite{Bergamin:2004us} dilaton theories in 2D. The absence of any local quantum effects suggests that these theories are fully equivalent to some effective quantum theories at the boundary.

\acknowledgments The authors thank Miguel Pino, Jakob Salzer  and Ricardo Troncoso for helpful conversations and comments. O.\,F. would like to thank Claudio Bunster and Alfredo P\'erez for useful discussions.  
M.\,C. acknowledges financial support given by Becas Chile, CONICYT.
 O.\,F. thanks the Institute for Theoretical Physics, TU Wien,  for the hospitality during his stay funded by the grant CONICYT PCI/REDES 170052.
The work of H.\,G. is supported by the Austrian
Science Fund (FWF), project P 28751-N2 and P 27182-N27.
C.\,V. was partially supported by CAPES post-doctoral scholarship. D.V. was supported by the grant 2016/03319-6 of the S\~ao Paulo Research Foundation (FAPESP),  by the grants 401180/2014-0 and 303807/2016-4 of CNPq, by the RFBR project 18-02-00149-a and by the Tomsk State University Competitiveness Improvement Program. This research has been partially supported by FONDECYT grant N 3170772.
The Centro de Estudios Cient\'ificos (CECs) is funded by the Chilean
Government through the Centers of Excellence Base Financing Program
of Conicyt.
This collaboration was initiated during the ESI programme \href{http://quark.itp.tuwien.ac.at/~grumil/ESI2017/}{``Quantum physics and gravity''} in Summer 2017.

\appendix
\section{Matrix representation of OSp\texorpdfstring{$\boldsymbol{(2,N)}$}{(2,N)}}\label{sec:A1}

The $\mathfrak{osp}(2,N)$ algebra is given by
\begin{gather}
\begin{align}
[J_{I},J_{J}]&=\epsilon_{IJK}J^{K} & [J_{I},Q_{\alpha}^{A}]& =\frac{1}{2}(\Gamma_{I})_{\,\,\,\alpha}^{\beta}Q_{\beta}^{A}\nonumber \\
[Q_{\alpha}^{A},T^{BC}]&=2\delta^{A[B}Q_{\alpha}^{C]} & [T^{AB},T^{CD}]&=4\delta^{[A|[C}T^{D]|B]}
\end{align}\\
\{Q_{\alpha}^{A},Q_{\beta}^{B}\}=-\frac{1}{2}\delta^{AB}\left(C\Gamma^{I}\right)_{\alpha\beta}J_{I}-\frac{1}{4}C_{\alpha\beta}T^{AB}\nonumber
\end{gather}
where capital letters $A,B,C,D,\dots$ run from $1$ to $N$. Here, the $\Gamma$-matrices satisfy the Clifford algebra in three dimensions; $\{\Gamma_{I},\Gamma_{J}\}=2\eta_{IJ}$,
where the metric is chosen as $\eta_{IJ}=\text{diag}(1,1,-1)$ and
$\epsilon_{012}=-1$. The generators of the $SO(N)$ subgroup were denoted as $T^{AB}$. The charge conjugation matrix is given by $C_{\alpha\beta}=\epsilon_{\alpha\beta}$, where $\epsilon_{+-}=-1$. 

The invariant bilinear metric associated to $\mathfrak{osp}(2,N)$  is given by the supertrace, whose nonvanishing components
are chosen as
\begin{equation}
\text{Str}[J_{I}J_{J}]=\frac{1}{2}\eta_{IJ}\qquad\text{Str}[Q_{\alpha}^{A}Q_{\beta}^{B}]=\frac{1}{2}C_{\alpha\beta}\delta^{AB}\qquad\text{Str}[T^{AB}T^{CD}]=4\delta^{A[C}\delta^{D]B}\label{eq:StrN}
\end{equation}
which is identically satisfied by the the spinorial representation
of  $\mathfrak{osp}(2,N)$;
\begin{equation}
J_{I}=\left(\begin{array}{c|c}
\frac{1}{2}(\Gamma_{I})_{\,\,\beta}^{\alpha} & \begin{array}{c}
0_{2\times N}\end{array}\\
\hline 0_{N\times2} & 0_{N\times N}
\end{array}\right)\quad Q_{\alpha}^{A}=\left(\begin{array}{c|c}
0_{2\times2} & \frac{1}{2}\delta_{\alpha}^{\beta}\delta_{C}^{A}\\
\hline -\frac{1}{2}\delta^{AB}C_{\alpha\gamma} & 0_{N\times N}
\end{array}\right)\quad T^{AB}=\left(\begin{array}{c|c}
0_{2\times2} & 0_{2\times N}\\
\hline 0_{N\times2} & 2\delta^{C[A}\delta_{D}^{B]}
\end{array}\right)\,.
\end{equation}
An explicit matrix representation for the $\Gamma$-matrices is given by
\begin{equation}
\Gamma_{0}=\left(\begin{array}{cc}
0 & 1\\
1 & 0
\end{array}\right)\qquad\quad\Gamma_{1}=\left(\begin{array}{cc}
1 & 0\\
0 & -1
\end{array}\right)\qquad\quad\Gamma_{2}=\left(\begin{array}{cc}
0 & -1\\
1 & 0
\end{array}\right)\,.
\end{equation}

In order to perform the analysis of the asymptotic structure of the theory it is
useful to express $\mathfrak{osp}(2,N)$ as follows
\begin{gather}
\begin{align}
\left[L_{m},L_{n}\right]&=\left(m-n\right)L_{m+n} & \left[L_{m},G_{p}^{A}\right]&=\left(\frac{m}{2}-p\right)G_{m+p}^{A}\\
[G_{p}^{A},T^{BC}]&=2\delta^{A[B}G_{p}^{C]} & [T^{AB},T^{CD}]&=4\delta^{[A|[C}T^{D]|B]}
\end{align}\\
\{G_{p}^{A},G_{q}^{B}\}=-2L_{p+q}\delta^{AB}-(p-q)T^{AB}
\end{gather}
where the map of the generators is given by
\begin{gather}
L_{0}=J_{1}\qquad\qquad L_{-1}=J_{2}-J_{0}\qquad\qquad L_{1}=J_{2}+J_{0}\\
G^{A}_{-\frac{1}{2}}=2Q^{A}_{+}\qquad\qquad G^{A}_{\frac{1}{2}}=2Q^{A}_{-}\,.
\end{gather}
The matrix representation in this basis reads
\begin{gather}
L_{-1}=\left(\begin{array}{c|c}
\begin{array}{cc}
0 & -1\\
0 & 0
\end{array} & 0_{2\times N}\\
\hline 0_{N\times2} & 0_{N\times N}
\end{array}\right)\quad L_{0}=\left(\begin{array}{c|c}
\begin{array}{cc}
\frac{1}{2} & 0\\
0 & -\frac{1}{2}
\end{array} & 0_{2\times N}\\
\hline 0_{N\times2} & 0_{N\times N}
\end{array}\right)\quad 
L_{1}=\left(\begin{array}{c|c}
\begin{array}{cc}
0 & 0\\
1 & 0
\end{array} & 0_{2\times N}\\
\hline 0_{N\times2} & 0_{N\times N}
\end{array}\right) \label{eq:Ls}\\
G^{A}_{-\frac{1}{2}}=\left(\begin{array}{c|c}
0_{2\times2} & \delta_{+}^{\beta}\delta_{C}^{A}\\
\hline -\delta^{AB}C_{+\gamma} & 0_{N\times N}
\end{array}\right)\quad G^{A}_{\frac{1}{2}}=\left(\begin{array}{c|c}
0_{2\times2} & \delta_{-}^{\beta}\delta_{C}^{A}\\
\hline -\delta^{AB}C_{-\gamma} & 0_{N\times N}
\end{array}\right)\,.\label{eq:Gs}
\end{gather}
The nonvanishing components of the supertrace are now given by
\begin{gather}
\text{Str}[L_{-1}L_{1}]=\text{Str}[L_{1}L_{-1}]=-1\qquad\qquad\text{Str}[L_{0}^{2}]=\frac{1}{2}\\
\text{Str}[G^{A}_{-\frac{1}{2}}G^{B}_{\frac{1}{2}}]=-\text{Str}[G^{A}_{\frac{1}{2}}G^{B}_{-\frac{1}{2}}]=-2\delta^{AB}\,.
\end{gather}

The group element in \eqref{gauss} and its inverse can be explicitly computed by using the representation \eqref{eq:Ls} and \eqref{eq:Gs}, which read
\begin{equation}
g=\left[\begin{array}{ccc}
e^{q} & -ke^{q} & e^{q}\zeta_{B}\\
he^{\frac{q}{2}} & e^{-q}-khe^{q}+M^{CD}\eta_{C}\zeta{}_{D} & e^{q}h\zeta_{B}+\eta^{C}M_{CB}\\
-e^{q}\eta^{A} & e^{q}k\eta^{A}+M^{AC}\zeta_{C} & -e^{q}\eta^{A}\zeta_{B}+M_{\,\,\,\,\,B}^{A}
\end{array}\right]
\end{equation}
and
\begin{equation}
g^{-1}=\left[\begin{array}{ccc}
e^{-q}-khe^{q}+M^{CD}\eta{}_{C}\zeta_{D} & ke^{q} & -e^{q}k\eta_{B}-M_{BC}\zeta^{C}\\
-he^{q} & e^{q} & -e^{q}\eta_{B}\\
e^{q}h\zeta^{A}+M^{CA}\eta_{C} & -e^{q}\zeta^{A} & e^{q}\zeta^{A}\eta_{B}+M_{B}^{\,\,\,\,A}
\end{array}\right]
\end{equation}
respectively.
\vspace{1cm}

\providecommand{\href}[2]{#2}\begingroup\raggedright

\bibliographystyle{fullsort}
\addcontentsline{toc}{section}{References}
\begingroup\raggedright

\end{document}